%% file: main.tex
\documentclass[acmsmall]{acmart}
\usepackage[utf8]{inputenc}

\PassOptionsToPackage{table,xcdraw}{xcolor} 

\usepackage{textgreek}
\usepackage{multirow}
\usepackage{subcaption}
\usepackage{xcolor,colortbl}
\definecolor{gray}{rgb}{0.1,0.1,0.1}
\usepackage[T1]{fontenc}
\usepackage{graphicx}
\usepackage{tabularx}
\usepackage{longtable}
\usepackage{enumitem}

\acmConference[Woodstock '18]{Woodstock '18: ACM Symposium on Neural
  Gaze Detection}{June 03--05, 2018}{Woodstock, NY}
\acmBooktitle{Woodstock '18: ACM Symposium on Neural Gaze Detection,
  June 03--05, 2018, Woodstock, NY}
\acmPrice{15.00}
\acmISBN{978-1-4503-XXXX-X/18/06}

\acmSubmissionID{123-A56-BU3}

\input{sections/header}

\begin{document}

\title[``A Party That Never Stops'': Designing Digital Third Places for Young People’s Friendships on Discord]{``A Party That Never Stops'': Designing Digital Third Places for Young People’s Friendships on Discord}

\author{JaeWon Kim}
\affiliation{%
  \institution{The Information School, University of Washington}
  \city{Seattle}
  \state{WA}
  \country{USA}}
\email{jaewonk@uw.edu}

\author{Thea Klein-Balajee}
\affiliation{%
  \institution{The Information School, University of Washington}
  \city{Seattle}
  \state{WA}
  \country{USA}}
\email{tkleinb@uw.edu}

\author{Ryan M. Kelly}
\affiliation{%
  \institution{RMIT University}
  \city{Melbourne}
  \state{VIC}
  \country{Australia}
}
\email{ryan.kelly@rmit.edu.au}

\author{Alexis Hiniker}
\affiliation{%
  \institution{The Information School, University of Washington}
  \city{Seattle}
  \state{WA}
  \country{USA}}
\email{alexisr@uw.edu}

\renewcommand{\shortauthors}{JaeWon Kim, et al.}

\begin{abstract}
Third places---informal settings where relationships grow through repeated casual contact---are increasingly inaccessible to young people. Prior work shows that virtual third places exist, but little is known about how persistent settings and control over participation work together as relationships form. We interviewed 25 active Discord users ages 15--24 who had reported forming at least one friendship on the platform. Their accounts highlighted two conditions that worked together. Persistent, socially legible servers let them return to the same community, recognize others, and build shared histories and norms. Controllable presence let them observe, participate, deepen a relationship, or withdraw while managing visibility, identity, and commitment. Graduated visibility was especially important for cautious entry and identity exploration. We translate these findings into design principles, showing how each can support some of Oldenburg's third-place functions while creating tensions with others. These findings suggest that persistent digital communities can complement increasingly scarce physical third places by giving young people recurring, low-pressure opportunities for social connection.
\end{abstract}

\ccsdesc[500]{Human-centered computing~Collaborative and social computing}

\keywords{Discord, online community, third place, identity, presence, pseudonymity, youth, friendship}

\maketitle

\input{sections/1_introduction}
\input{sections/2_related-work}
\input{sections/3_method}
\input{sections/4_results}
\input{sections/5_discussion}
\input{sections/6_conclusion}

\begin{acks}
The authors thank the University of Washington Global Innovation Fund (GIF) and Student Technology Fee (STF) for supporting this work. We are grateful to the anonymous reviewers for their detailed feedback, to our participants for generously sharing their perspectives, and to Daniel Enriquez and Lindsay Popowski for their valuable comments on earlier drafts. Alexis Hiniker is a special government employee for the Federal Trade Commission; the content expressed in this manuscript does not reflect the views of the Commission or any of the Commissioners.
\end{acks}

\bibliographystyle{ACM-Reference-Format}
\bibliography{references}

\input{sections/7_appendix}

\end{document}

%% file: sections/header.tex
\input{formatting/published-packages} 
\input{formatting/published-commands} 


%% file: formatting/published-packages.tex
\usepackage{xspace}     
\usepackage{xpunctuate} 

%% file: formatting/published-commands.tex
\newcommand{\inlinequote}[1]{\emph{``#1''}\allowbreak} 

\newcommand{\parHeading}[1]{\vspace{2mm}\noindent{\textbf{#1.}\allowbreak}}

\newcommand{\cut}[1]{}

%% file: sections/1_introduction.tex
\section{Introduction}
\label{sec:introduction}

\begin{quote}\emph{From wealthy suburbs to small towns, teenagers reported that parental fear, lack of transportation, and structured lives restricted their ability to meet friends face to face\ldots{} At home, teens grappled with lurking parents. The formal activities teens described were so structured that they allowed little room for casual sociality\ldots{} This prompts teens to desperately---and, in some cases, sneakily---seek it out. As a result, many turn to asynchronous social media, texting, and other mediated interactions.} (boyd, 2014, p. 90) ~\cite{BoydDanah2014ICTS}\end{quote}

Human connection develops both online and in what Oldenburg called ``third places,'' or ``informal public gathering places''~\cite{Oldenburg1997-fb}. Third places support collective action~\cite{Putnam2000-bowling, Horrigan2001-wg} and, more fundamentally, community and interpersonal connection~\cite{Cattell2008-wm, Bosman2019-dq, Oldenburg1999-nd}. Yet public places for casual gatherings are declining~\cite{Klinenberg2018-palaces, Rao2024-mj, Oldenburg1997-fb}. This loss especially affects young people, whose access often depends on transportation, money, and adult permission~\cite{BoydDanah2014ICTS, Skelton2000-nothing}. Because our participants were 15 to 24---ages when peer relationships and identity development are especially important~\cite{erikson_identity_1994, arnett2000emerging}---the design of digital third places bears directly on their social development.

Research shows that virtual third places exist and identifies mechanisms that sustain sociability in online games, streaming communities, and social worlds~\cite{Steinkuehler2006-tm, Ducheneaut2007-kv, Hamilton2014-vf, Moore2009-qg}. Recent studies also identify Discord as a third place~\cite{vanOmmen2025-belonging, Ringeis2026-beyond}. Other work examines placemaking~\cite{DourishHarrison-1996-Re-place-ingSpaceSystems-f}, presence~\cite{Erickson2000-gn, Licoppe2004-connected}, pseudonymity and identity~\cite{Marwick2011-lr, Ellison2016-yp}, lurking~\cite{Nonnecke2000-lurker}, and online relationship development~\cite{Sheng2020-ym}. What remains unclear is how returnable settings and adjustable participation work \textit{together} as relationships form, including the design tensions this combination creates.

Discord is a useful site for this question. In prior interviews on another topic, teens spontaneously identified it as a place where friendships formed and comfortable self-disclosure occurred (\S\ref{section:method}). Discord combines persistent communities, multimodal communication, contextual identities, and low-obligation entry (\S\ref{sec:background}) across gaming, friend, fandom, and educational servers~\cite{Kelly2021-yl, Hull2021-wh, Browning2021-ki, Lorenz2019-bg, Soeiro2023-ss, Vladoiu2020-yj}. Because participants reported third-place experiences without embodied avatars or a navigable world, Discord offers a focused case for examining what matters when simulated realism is absent.

We therefore ask:
\begin{itemize}
    \item \textbf{RQ1:} How do young people describe forming friendships and developing a sense of community belonging on Discord?
    \item \textbf{RQ2:} What platform and community arrangements do they associate with entering, sustaining, and deepening those relationships?
\end{itemize}

Across 25 semi-structured interviews, participants described persistent, socially legible settings (\S\ref{sec:returnable-settings}) and controllable, reversible participation (\S\ref{sec:controllable-participation}) working together. Repeated low-obligation encounters built familiarity, after which participants could deepen ties through reversible steps. Participants associated this pattern with friendships, romantic relationships, and community belonging (\S\ref{sec:results-outcomes} and \S\ref{sec:synthesis}), but most encounters remained shallow. We also report cases in which relationships stalled, reversed course, or followed a different route.

This paper contributes: (1) an empirical account of this pathway and where it breaks down; (2) the concept of \textit{controllable presence}, including graduated visibility and its safety implications; and (3) design principles for platforms and communities, showing how each can support several of Oldenburg's third-place functions while creating tensions with others. Together, these findings suggest that persistent digital communities can complement increasingly scarce physical third places by giving young people more opportunities for low-pressure, recurring social connection.

%% file: sections/2_related-work.tex
\section{Related Work}
\label{sec:related-work}
\subsection{Understanding ``Third Places'' in Physical and Virtual Space}
\label{sec:third-places}
Ray Oldenburg~\cite{oldenburg1982third, Oldenburg1999-nd} defines ``third places'' as inclusive spaces outside the home (the first place) and work (the second place), where relationships develop through repeated informal interactions. Everyday gathering spots such as cafes, libraries, parks, and bars anchor community life outside the structured demands of family and employment.

According to Oldenburg, such spaces exhibit eight key characteristics: 1) ``on neutral ground'' where people can come and go freely; 2) social ``leveling,'' making status irrelevant; 3) conversation as the main activity; 4) accessibility and accommodation; 5) regular visitors shaping the ambiance and welcoming newcomers; 6) a ``low profile'' that avoids pretension; 7) a ``playful mood'' that encourages humor and wit; and 8) a sense of being a ``home away from home.'' These qualities highlight how seemingly ordinary environments can sustain strong social bonds and local community life.

Oldenburg attributed third-place conviviality to casual drop-ins, repeated encounters, and shared physical presence. Yet his account emphasizes proximity without establishing how important it is. Platforms such as Discord and Twitch create co-presence without physical closeness, allowing researchers to ask which properties actually sustain third-place experiences.

Scholars have long explored virtual spaces as digital counterparts to physical third places~\cite{Soukup2006-sb, Kendall2002-xn, Foster2013-kw, Markiewicz2019-wa}. Drawing on Goffman's view of interaction as performance~\cite{goffman2016presentation}, Soukup argues that virtual place is ``self-selected and mutually constructed'' through repeated interaction around shared interests~\cite{Soukup2006-sb}. Chatrooms and other online spaces likewise foster playful conversation among emerging regulars~\cite{Parks1998-hr, Baym1995-ue, Schuler1996-la, Kim2015-bi, McArthur2016-mb, Alcala2023-zy}. Several studies examine how game structures produce sociability~\cite{Steinkuehler2006-tm}, user-created spaces and roles organize encounters~\cite{Ducheneaut2007-kv}, and hosts manage entry and group size~\cite{Moore2009-qg}. This work explains how virtual settings become sociable, but not how settings and participation work together as relationships form.

Recent work identifies Discord itself as a third place. Van Ommen and Yahanashi describe a Japanese competitive-gaming server that fostered belonging and emotional support~\cite{vanOmmen2025-belonging}; Ringeis and colleagues show how pseudonymity helps distinguish Discord servers from home and work~\cite{Ringeis2026-beyond}; and Sheng and Kairam find that ties begun on Twitch deepen in associated Discord servers~\cite{Sheng2020-ym}. These studies motivate our account of how place and participation interact and what design tensions arise.

\subsection{Virtual Platforms and the Role of Space}
\label{sec:virtual-platforms-space}
Spatial metaphors such as \inlinequote{virtual architecture,} \inlinequote{cyberspace,} \inlinequote{gateway,} and \inlinequote{chat rooms} can help users understand and navigate digital environments~\cite{Adams2010-io, Giese1998-gp, Maglio1999-il, kim2026metaphors}. By drawing on familiar spatial reasoning, they can aid memory and reduce cognitive effort~\cite{Kuhn1996-ro}.

Research on physical proximity shows that spatial arrangements shape repeated encounters and, therefore, opportunities for friendship~\cite{festinger1950social}. Networked communication has loosened but not severed the link between community and shared location~\cite{Rainie2012-gm}. Small and Adler identify three ways space can foster relationships: bringing people close together, providing fixed places for interaction, and using barriers or pathways to organize movement~\cite{Small2019-cb}. Digital platforms create similar conditions without requiring people to share a physical location. We ask how these conditions interact with users' control over participation.

\subsection{How Virtual Spaces Become Places}
\label{sec:virtual-spaces-places}
HCI research distinguishes the structure of digital space from the culture of place~\cite{DourishHarrison-1996-Re-place-ingSpaceSystems-f}. Harrison and Dourish summarize: \inlinequote{Space is the opportunity; place is the understood reality.} Place emerges as inhabitants adapt and appropriate a space, developing shared norms and meanings. Related work describes virtual places as distinct, recognizable, bounded environments with behavioral expectations~\cite{Freeman2012-br}; shows how repeated interaction creates presence and immersion~\cite{Saunders2011-mp}; and identifies identity, agency, and traces of memory as sources of placeness~\cite{Nitsche2009-vn}. Together, these accounts define place as socially produced rather than reducible to spatial structure.

A Discord server therefore provides infrastructure for a place but does not automatically become one. Place depends on repeated use, shared history, norms, governance, and awareness of others. Shared interests or activities---what Feld calls \textit{foci}---can generate ties~\cite{Feld1981-focused}, while a sense of virtual community arises through member behavior rather than software alone~\cite{Blanchard2004-experienced}. Presence likewise develops through visible activity~\cite{Erickson2000-gn}, frequent small exchanges~\cite{Licoppe2004-connected}, and ambient co-presence, or awareness of others without active conversation~\cite{Ito2005-technosocial}. Our participants' settings combined these elements: shared foci, legible activity, and presence below the level of active conversation.

Digital settings become socially legible through roles, shared times, and partitions~\cite{Ducheneaut2007-kv, Soukup2006-sb}; shared foci~\cite{Maglio1999-il, Bruckman1995-ge}; active hosts~\cite{Moore2009-qg}; visibility, awareness, and accountability~\cite{Erickson2000-gn, kim2026designspace}; and clear boundaries~\cite{Soukup2006-sb, Zhang2013-es, Maglio1999-il}. This literature explains how settings become places. A separate body of work examines how people control their participation within them.

Social information processing theory holds that repeated text exchanges can accumulate into relationships comparable in depth to face-to-face ties~\cite{Walther1992-sip}, while the hyperpersonal model explains how selective self-presentation can intensify mediated connection~\cite{Walther1996-hyperpersonal}. Modality-switching research examines movement between channels and into offline contact~\cite{Ramirez2007-when}. The third-place lens adds the persistent setting in which these interactions accumulate.

Identity and disclosure research shows how people manage audiences and self-presentations~\cite{kim2024bereal, kim2025trust}, including by maintaining boundaries across sites~\cite{Stutzman2012-ev}, using anonymity for sensitive conversations~\cite{Ellison2016-yp, Andalibi2018-by}, creating secondary accounts~\cite{Tao2023-an}, and rebuilding networks during identity transitions~\cite{Haimson2016-zr}. These practices respond to context collapse~\cite{Marwick2011-lr, Vitak2012-ne} and the pressure of persistent, real-name profiles~\cite{goffman2016presentation, Turkle1995-life}. Other research shows that newcomers learn from a community's edges~\cite{Lave1991-situated} and that lurking is often deliberate rather than free-riding~\cite{Nonnecke2000-lurker}. Our participants were also managing how much of themselves others could see: how observable, identifiable, and committed they were within a persistent community. We call this broader control \textit{controllable presence}, and we call its observability dimension \textit{graduated visibility} (\S\ref{sec:controllable-presence-discussion}). These concepts differ from prior work focused on others' visibility, profile-based self-presentation, community learning, or audience collapse~\cite{Erickson2000-gn, Barta2024-visibility, Nonnecke2000-lurker, Marwick2011-lr}.

Control over exposure also affects safety. Adolescent online-safety research advocates designs that preserve autonomy while managing risk~\cite{Wisniewski2018-rc, razi2020privacy, Agha2023-mu, kim2025privacy}, while justice-oriented work links governance to how harm is addressed~\cite{Schoenebeck2021-oh}. On Discord, teens describe vigilance during risky interactions~\cite{Koung2026-teen} and collective privacy sensemaking~\cite{Dong2026-collective}. We therefore examine beneficial arrangements alongside the harms they enable.

\section{Background: Discord as a Platform}
\label{sec:background}
Discord combines real-time voice, video, and text within persistent community spaces~\cite{discord_company}. Launched in 2015 as a lightweight voice tool for gamers, it now serves educational, professional, and social communities, while retaining features such as screen sharing, live streaming, and game-activity displays. Discord reached 200M monthly active users in 2024~\cite{Business-of-Apps2020-fb}; 28\% of surveyed young people reported using it~\cite{Anderson2023-qk}. Because the platform changes continuously, we describe features available during our July--September 2024 interviews. The interface mock-ups (Figure~\ref{fig:screenshot-1} and Appendix~\ref{appendix:platform}) were produced in September 2024.

\input{inserts/screenshots/screenshot-1}

\parHeading{Servers, channels, and categories}
The fundamental organizational unit is the \textit{server}: a persistent virtual space, joined via invitation or search, that contains text and voice \textit{channels} grouped under labeled categories (Figure~\ref{fig:screenshot-1}(a))~\cite{discord_beginner}. Users belong to many servers at once, and channels can be public to all members or restricted to holders of particular permissions.

\parHeading{Roles, permissions, and visible hierarchy}
Role-based permissions govern what each member can do, from sending messages to banning users~\cite{discord_roles}. Roles are hierarchical and visible. A member's highest role sets their username color, and servers can configure selected roles to appear as separate groups in the member list, allowing the list to reflect role and status regardless of participation (Figure~\ref{fig:screenshot-1}(d)). Members can also self-assign roles associated with interest groups (Figure~\ref{fig:screenshot-1}(c)). Communities therefore have visible local hierarchies: administrators and moderators hold publicly displayed positions with the power to mute, kick, or ban users; operators assign roles at their discretion; and paid subscriptions (Nitro) confer visible markers such as profile badges~\cite{discord_nitro}. This structure matters for our later discussion of Oldenburg's leveler characteristic (\S\ref{sec:identity-flexibility} and \S\ref{sec:design-principles}).

\parHeading{Entry gates, preview, and graduated entry}
Servers commonly gate full access behind rules screening or bot-run verification, admitting newcomers to a restricted state until they complete it (Figure~\ref{fig:screenshot-1}(b))~\cite{discord_rules_screening}. Server Discovery, meanwhile, lets users browse public servers, and its Preview Mode lets them read a server's public channels without joining. Users can therefore observe a community before making any commitment~\cite{discord_discovery}.

\parHeading{Voice presence and drop-in interaction}
Voice channels use a drop-in, drop-out model. They do not ring recipients; users click to join, can see who is present before entering, and can observe when others join or leave~\cite{discord_beginner}. Voice channels support screen sharing and synchronized viewing, and each has its own text chat.

\parHeading{Profiles, presence cues, and notification control}
Profiles show a chosen display name, avatar, bio, optional pronouns, mutual friends and servers, and current activity, such as the game being played and for how long. Nicknames can differ across servers, and paid subscribers can also set server-specific avatars, banners, and bios~\cite{discord_nitro}. Every user has a visible status (online, idle, do not disturb, or invisible), and notifications can be adjusted or muted by server, channel, and person~\cite{discord_notifications}. Users can therefore adjust their own reachability.

\parHeading{Custom expression, bots, and activities}
Servers upload their own emoji, which appear in messages and reactions and often become community inside jokes~\cite{discord_emoji}; third-party bots automate moderation, music playback, and recurring prompts~\cite{discord_developer}; and built-in Activities allow users to play and watch content together inside voice channels~\cite{discord_activities}.

Appendix~\ref{appendix:platform} provides an extended description of these features, with additional annotated interface mock-ups.

%% file: inserts/screenshots/screenshot-1.tex
\begin{figure}[htbp]
    \centering
    \includegraphics[width=0.95\linewidth]{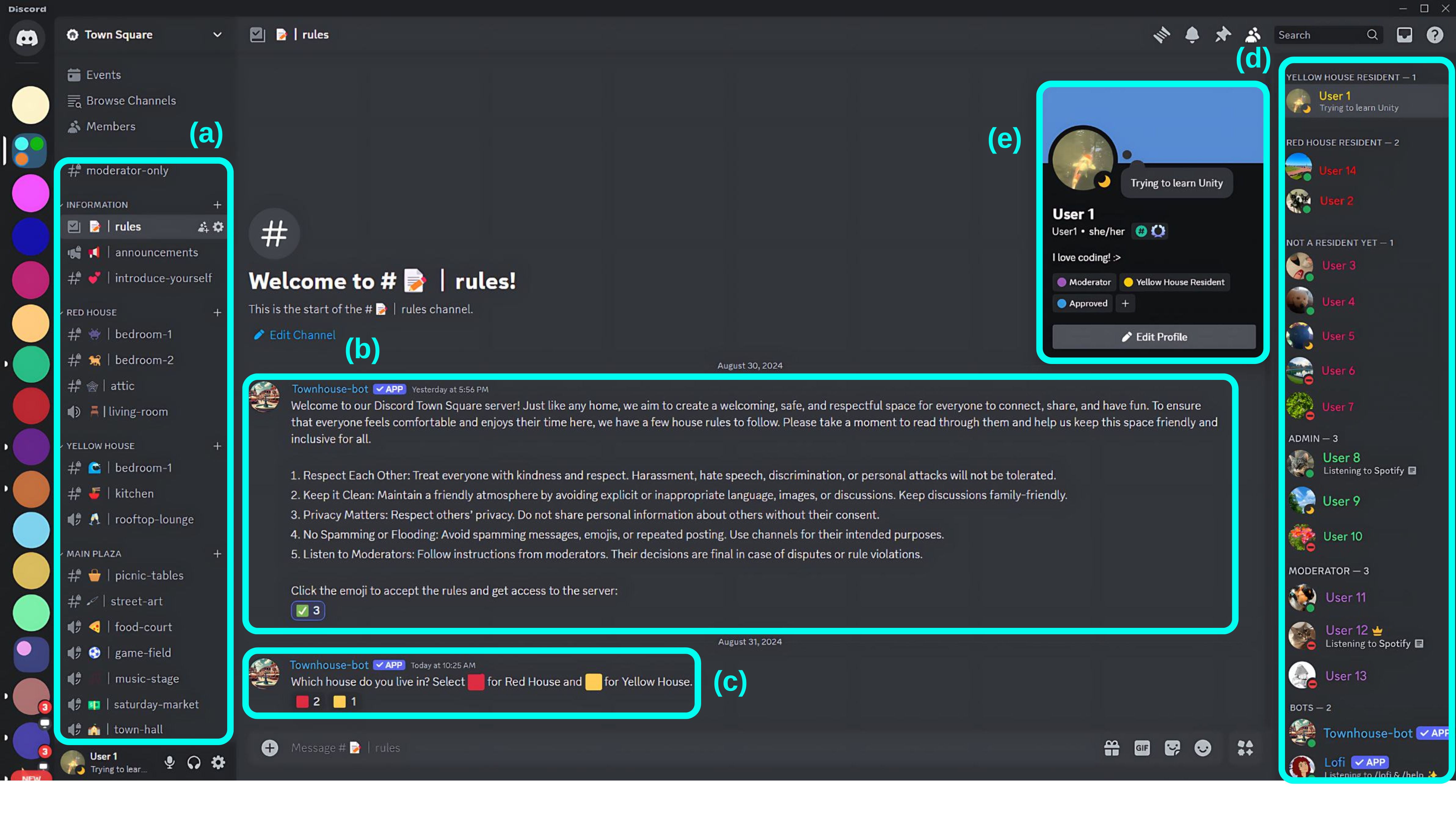}
    \caption{Annotated mock-ups of Discord rules and roles: (a) channel categories; (b) bot-managed entry rules, shown after acceptance; (c) self-assigned roles that control channel access; (d) member list organized by role; and (e) profile with a bio, roles, current activity, and a server-specific nickname.}
    \Description {This image is a mock screenshot of a Discord server interface with five highlighted sections for different functions. Section labeled (a): there is a vertical panel that displays the list of channels organized into different categories, including ``INFORMATION,'' ``RED HOUSE,'' ``YELLOW HOUSE,'' and ``MAIN PLAZA.'' Channels like ``rules,'' ``announcements,'' ``introduce-yourself,'' ``bedroom-1,'' and ``living-room'' are visible. Each channel appears with a distinct icon representing its purpose; Section labeled (b): In the main chat window, there is a message from ``Townhouse-bot'' outlining the server's rules such as maintaining privacy, avoiding spam, and following instructions from moderators. Below the rules, there is an interactive prompt where users can click an emoji to accept the rules and gain access to the server; Section labeled (c): Further down in the chat, there is another message from ``Townhouse-bot'' inviting users to select a house to join by reacting with specific emojis. The red square emoji represents the ``Red House,'' while the yellow square emoji represents the ``Yellow House,'' allowing users to make a choice by reacting accordingly; Section labeled (d): On the right side of the screen, a vertical panel shows the list of members in the server, categorized by roles such as ``YELLOW HOUSE RESIDENT,'' ``RED HOUSE RESIDENT,'' ``NOT A RESIDENT YET,'' ``ADMIN,'' ``MODERATOR,'' and ``BOTS.'' Members' usernames, avatars, and statuses are visible, with some indicating activities like ``Listening to Spotify'' or personal statuses like ``Trying to learn Unity''; Section labeled (e): To the right of the chat window, a detailed user profile is displayed for ``User 1,'' showing information such as their role as ``Moderator'' and ``Yellow House Resident,'' their pronouns ``she/her,'' and a personal status message that reads ``Trying to learn Unity.'' Badges or icons indicating roles like ``Moderator'' and ``Approved'' are also shown in the profile.}
    \label{fig:screenshot-1}
\end{figure}

%% file: sections/3_method.tex
\section{Method}
\label{section:method}

We conducted semi-structured interviews with 25 young Discord users about forming friendships and community belonging and the arrangements they associated with those relationships. We recruited people who had reported forming a friendship on Discord so we could examine how those relationships developed in detail. This sample cannot show how often Discord use leads to friendship, compare platforms, estimate harm, or represent users whose primary experience was failed connection or harm.

\subsection{Platform Selection and Author Positionality}
\label{sec:platform-selection}
Our choice of Discord followed prior interviews on another topic, in which 7 of 19 teens spontaneously identified it as a place for friendship or comfortable self-disclosure (\S\ref{sec:introduction}). This pattern and our interest in spatial context motivated the present study.

The first and last authors, who led the initial design, were not Discord users; the second and third were. This insider--outsider mix helped the team question assumptions and correct misunderstandings about platform practices and community norms during protocol design and coding. All authors shared an interest in spatial and third-place framings, which shaped both the study questions and our deductive use of Oldenburg's characteristics (\S\ref{sec:data-analysis}).

\subsection{Participants and Recruitment}
\label{sec:participants}
We used purposive and convenience sampling through a screener distributed to a consented contact list from prior studies and through campus flyers. Of 74 respondents, 55 were eligible; we interviewed the first 25 available and compensated each with a \$25 Amazon gift card. First-come scheduling favored people with immediate availability, and both recruitment channels tied the sample to one university. Recruitment ran from June to July 2024 and interviews from July to September 2024. Our institution's IRB approved the protocol. Adults consented; for minors, a parent or guardian provided written consent and the participant assented.

Eligibility required being 15--24 years old, using Discord for at least 15 minutes weekly for three months, and reporting that they had formed a friendship there. This range spans later adolescence, the United Nations youth category~\cite{un-youth}, and emerging adulthood~\cite{arnett2000emerging}, when peer relationships carry particular weight~\cite{erikson_identity_1994, arnett2000emerging}. We treat these ages as heterogeneous: 10 participants were minors and 15 were ages 18--24 (Appendix~\ref{appendix:demographics}).

Participants' mean age was 18.0 (N=25, min=15, max=24, SD=1.97), and their mean Discord use was 46.3 months. Gender options were not mutually exclusive: 17 participants (68\%) selected Girl/Woman, seven (28\%) Boy/Man, and three (12\%) Non-binary or Third Gender; P19 and P22 selected both Girl/Woman and Non-binary. Eighty percent reported friendships beyond casual acquaintance, while five (20\%) later described their Discord connections as casual. Because the screener did not define ``friendship,'' eligibility was effectively self-defined. Table~\ref{tab:demographics} and Appendix~\ref{appendix:demographics} provide full demographics.

\input{inserts/demographics-summary}

\subsection{Procedure}
\label{sec:procedure}

The first author conducted 21 interviews and the second conducted four. Each 60-minute Zoom interview covered participants' use and perceptions of Discord, relationships formed there, and features that helped or hindered relationship building, including comparisons with other platforms. Appendix~\ref{appendix:protocol} provides the protocol.

When participants described Discord as effective for friendship formation, we asked them to compare it with structurally similar platforms such as Facebook Groups and Reddit. Hypothetical prompts separated interface structure from demographics or content quality; for example, \textit{``If Facebook Groups also had mostly younger users and less spam, which would you prefer for making new connections?''} These were not controlled comparisons, so we treat the responses as participants' interpretations. The protocol never mentioned ``third places'' or Oldenburg's framework. It also asked about barriers, disliked features, and each platform's metaphorical ``vibe'' (Appendix~\ref{appendix:protocol}).

We distinguish spatial analogies that arose spontaneously from those prompted by the ``vibe'' question. Although the protocol did not introduce our later framing, its focus on Discord and on friendships participants had already formed still shaped the data.

\subsection{Data Analysis}
\label{sec:data-analysis}
We combined deductive codes based on Oldenburg's eight characteristics~\cite{Oldenburg1999-nd} with inductive codes for material outside that framework. Inductive coding retained themes such as ambient use (\S\ref{sec:background-use}) and stalled or dissolved relationships (\S\ref{sec:synthesis}). We analyzed 25 Zoom-generated transcripts.\footnote{Coders checked audio whenever a transcript passage was garbled or ambiguous but did not systematically re-check every recording.}

All four authors independently coded two transcripts line by line, applying deductive codes and generating inductive ones. The first author refined the codebook; the first and second authors then coded four more transcripts, and the first author coded the remaining 19. The team discussed ambiguous or ill-fitting segments and resolved disagreements by consensus. We do not report inter-rater reliability because consensus coding and reliability coefficients serve different analytic goals~\cite{McDonald2019-tu}. We used Atlas.ti~\cite{atlas}.

Because Oldenburg's characteristics did not map directly onto design decisions, we first grouped codes by what an arrangement did. We then organized them into two themes: making settings persistent, returnable, and socially legible; and making participation controllable, reversible, and low-obligation. Supplementary materials provide definitions, note whether each code was deductive or inductive, and include de-identified examples and participant coverage.

%% file: inserts/demographics-summary.tex

\begin{table}[h]
\small
\centering
\renewcommand{\arraystretch}{1.2}

\caption{Participant demographics (N=25).}
\label{tab:demographics}
\begin{tabular}{p{4cm} p{9.5cm}}
\toprule
\textbf{Characteristic} & \textbf{Participants} \\
\midrule

{Gender identity\textsuperscript{a}} & Girl/Woman (17; 68\%), Boy/Man (7; 28\%), Non-binary or Third Gender (3; 12\%) \\

{Age} & 15 (2; 8\%), 16 (4; 16\%), 17 (4; 16\%), 18 (4; 16\%), 19 (7; 28\%), 20 (3; 12\%), 24 (1; 4\%) \\

{Race} & Asian or Asian-American (14; 56\%), White (8; 32\%), Black or African-American (2; 8\%), Hispanic or Latino (1; 4\%) \\

{Social media platforms used} & Discord (25; 100\%), Instagram (23; 92\%), TikTok (18; 72\%), Snapchat (17; 68\%), Reddit (15; 60\%), Facebook (13; 52\%), Twitter/X (11; 44\%), BeReal (11; 44\%), Threads (5; 20\%), LinkedIn (1; 4\%), GroupMe (1; 4\%), Slack (1; 4\%), Noplace (1; 4\%) \\

{Frequency of new Discord connections} & \textit{2: Rarely} (I have made 1-2 new connections, but it’s not common for me.) (1; 4\%); \textit{3: Occasionally} (I make new connections from time to time, but it’s not frequent.) (13; 52\%); \textit{4: Frequently} (I regularly make new connections on Discord.) (9; 36\%); \textit{5: Very Often} (I make new connections on Discord almost all the time.) (2; 8\%) \\

{Discord connection experience} & \textit{I've formed close, lasting friendships on Discord} (11; 44\%), \textit{I've made several good friends} (9; 36\%), \textit{I've made a few casual acquaintances} (5; 20\%) \\

\bottomrule
\end{tabular}

\smallskip
\noindent{\small \textsuperscript{a}\,Participants could select more than one gender identity. P19 and P22 selected both Girl/Woman and Non-binary or Third Gender, so the counts total 27 and the percentages total 108\%.}
\end{table}


%% file: sections/4_results.tex
\section{Results}
\label{sec:results}

\subsection{Discord as a Returnable Social Setting for Friendship and Belonging}
\label{sec:results-outcomes}
Participants described three outcomes that the rest of this analysis explains: recognizable places, relationships they called close, and communities to which they felt they belonged. The arrangements they associated with these outcomes follow in \S\ref{sec:returnable-settings} and \S\ref{sec:controllable-participation}.

\subsubsection{Participants Described Discord Through Place-Like Analogies}
\label{sec:place-analogies}
Participants described Discord in spatial terms, sometimes spontaneously and sometimes in response to a prompt inviting metaphorical descriptions of platforms (\S\ref{sec:procedure}, Appendix~\ref{appendix:protocol}). They compared Discord to well-defined places with distinct social functions. P11 likened it to a \inlinequote{town square} for gathering and branching into niche areas, while P20 saw it as a \inlinequote{concert hall} with themed rooms where returning visitors \inlinequote{immediately notice a couple of people, you know}. P07 compared it to a \inlinequote{cafe} where strangers communicate like friends, and P16 described it as a \inlinequote{hotel} with spaces for discussions and private conversations. P24 described a server structured like a \inlinequote{house} with rooms serving different functions. These descriptions often emerged in contrast with platforms like Instagram, which participants felt lacked a comparable sense of place. The analogies share several properties: bounded settings, distinguishable rooms, and people who can be found there again. We call such a configuration a \textit{returnable setting}: a bounded social space in which people, prior interactions, local norms, and shared history can be encountered again.

\subsubsection{Repeated Interaction Supported Close Relationships}
\label{sec:close-relationships}
Participants described close relationships that formed through Discord and distinguished them from more casual contacts. P07 described some contacts as \inlinequote{not really close with\ldots{} only the server}, while calling a Discord-made friend with whom he exchanged relationship advice and now shares locations \inlinequote{like best friends now, essentially}. P22 described friends made through servers dedicated to \inlinequote{very niche hobbies}, noting that they \inlinequote{only knew [them] over Discord} at first but now talk regularly and have met in person \inlinequote{literally dozens and dozens of times}. P19 explained how a school server's math-help channel led to lasting ties: \inlinequote{it would start off as\ldots{} math help\ldots{} which then turned into daily texting, and\ldots{} some of the friendships I still have.} P09, P19, and P22 described romantic relationships that began with someone they first encountered on Discord; P10 described one that lasted five months before contact ended. These accounts echo Tidwell's description of mediated connection as \inlinequote{not in-person but not impersonal}~\cite{Tidwell2023-bt}.

\subsubsection{Shared Settings Supported Community Belonging}
\label{sec:community-belonging}
Participants described belonging as attachment to the group and seeing themselves as part of a ``we''~\cite{Blanchard2004-experienced}. P11 explained that seeing \inlinequote{the same people very frequently} created a \inlinequote{community aspect} where they felt \inlinequote{a sense of joy and\ldots{} belonging} and found it \inlinequote{easier to make genuine friends.} P07 described the social climate this attachment produced: \inlinequote{once you're in the community, everybody just wants to make friends\ldots{} everyone's friendly.}

Participants also described their networks growing through these communities. P24 appreciated finding people \inlinequote{that aren't necessarily very easy to find\ldots{} in person} through niche communities. P01 described how her network \inlinequote{just basically grew\ldots{} friends of friends of friends}, including one close friend she is \inlinequote{still friends\ldots{} today\ldots{} which is crazy} with, years after they met. These accounts emphasize multiplying contacts, while the accounts above emphasize attachment to the group itself.

The rest of the section examines the arrangements participants associated with these outcomes and how those arrangements worked together in individual accounts.

\subsection{Making Settings Persistent, Returnable, and Socially Legible}
\label{sec:returnable-settings}
Four arrangements made servers recognizable and returnable in participants' accounts: persistence, channel segmentation, visibility cues, and community configuration. Together, they supported repeated encounters and made other users recognizable, while also creating tensions among third-place functions.

\subsubsection{\texorpdfstring{Persistent Spaces Supported Repeated Encounters}{Persistent Spaces Supported Repeated Encounters}}
\label{sec:persistent-spaces}
Persistence preserved shared history and made repeated encounters possible: a server remained findable, its people recognizable, and its prior conversations recoverable. P16 likened this permanence to \inlinequote{a party that never stops,} appreciating the ability to \inlinequote{leave, and\ldots{} come back,} knowing \inlinequote{it's always gonna be there.} Participants connected this stability to the repeated contact through which acquaintances became friends. P12 explained: \inlinequote{Discord has a much stronger way to make friends\ldots{} Instagram is more random\ldots{} but on Discord, you're in the same channels, see each other regularly\ldots{} chat, come back a day later, and cross paths again. There's not nearly that much on Instagram.} Seeing \inlinequote{the same name over and over} (P11) helped participants distinguish who was \inlinequote{in-group} and \inlinequote{out-group} (P11), and repeated encounters sometimes led to \inlinequote{a small connection} (P06). P06 added that public-server connections could \inlinequote{die out really quickly} when nothing more personal followed, a pattern we return to in \S\ref{sec:synthesis}.

Persistence supported the formation of regulars and a home away from home, but it could also create obligation. P01 checked Discord partly because her boss sent work updates there. She also described servers she had \inlinequote{just joined\ldots{} and then I forget to leave them. So it just builds up}, experiencing persistence as clutter and expectation.

\subsubsection{\texorpdfstring{Channels Organized Group Size and Conversation}{Channels Organized Group Size and Conversation}}
\label{sec:channel-segmentation}
P21 explained how participants used channels to manage group size, topics, and boundaries without leaving the community: \inlinequote{People just choose not to [all join or be active in the same channels]\ldots{} It's kind of like how a house has multiple rooms. If you're walking through your house and you see your sister talking on the phone to someone in her room, you're not going to go in that room. You're gonna find another room.}

P12 said that in \inlinequote{larger servers, people get buried away\ldots{} it's harder to see the same people over and over.} P02 likewise described smaller group chats as \inlinequote{more intimate than just a server.} Channels also allowed different topics to unfold in parallel. P25 valued having \inlinequote{a general channel [and] an off-topic channel\ldots{} you end up getting to know people a lot better through things like that.} P10 observed that users could \inlinequote{move [conversations to] different places while keeping the original conversation going} when they became \inlinequote{sidetracked}. On texting platforms, P10 said, users could address new topics immediately and \inlinequote{lose the previous topic} or \inlinequote{risk forgetting [the new topics]}. Channel segmentation supported conversation, accessibility, and intimate subspaces within a larger community. However, P24 found the platform's flexible structure overwhelming, and smaller spaces could develop insider groups that made entry more difficult.

Segmentation also helped participants manage several conversations. P19 said switching between servers and channels was \inlinequote{really easy}, allowing them to manage \inlinequote{2 or 3 conversations at once} or even \inlinequote{4 conversations with 2 people}. P07 preferred Discord to Instagram for relationship-building because \inlinequote{you can talk to multiple people at a time} and \inlinequote{it's just way faster.} P10 noted that Instagram required \inlinequote{constantly clicking in and out}, while Discord let users have \inlinequote{your quotes channel pulled up and your main chatting channel pulled up} at the same time. P06 used channels and pinned messages to \inlinequote{make plans in an organized way}, calling Discord \inlinequote{a really efficient way to socialize with new people.} P04 contrasted this structure with Facebook groups, which \inlinequote{don't have as much categorization capabilities}: what would be ten channels in one Discord server becomes \inlinequote{10 different Facebook groups\ldots{} I can't find them under one roof.}

\subsubsection{\texorpdfstring{Visibility Cues Showed Who Was Present and Available}{Visibility Cues Showed Who Was Present and Available}}
\label{sec:visibility-cues}
Visibility cues showed who was present, what they were doing, and whether they seemed available. In face-to-face settings, people perceive these signals directly; virtual spaces require explicit cues to create similar social translucence~\cite{Erickson2000-gn}. Discord's status indicators showed \inlinequote{what song [they're] listening to} (P21) or \inlinequote{which game they're playing and how long} (P07). Participants used this information to decide whether to \inlinequote{not bother} someone, \inlinequote{ask questions about the game} (P21), or \inlinequote{check in} (P07). Seeing active friends sometimes prompted P20 to think, \inlinequote{`maybe I'll go join them'}. Participants also valued features showing \inlinequote{friends that are active right now} (P20). P05 compared join notifications to \inlinequote{opening a door to conversation}: a casual \inlinequote{`Hi!'} signaled availability without demanding a response.

These cues supported recognition, conversation, and low-cost entry. They also exposed when someone was online and what they were doing, creating privacy concerns and pressure to appear available. Participants' control over these cues appears in \S\ref{sec:controllable-participation}.

\subsubsection{\texorpdfstring{Communities Shaped Their Own Structure and Norms}{Communities Shaped Their Own Structure and Norms}}
\label{sec:community-control}
Community control let members shape their own settings through server names, channel structures, roles, and rules. Participants used these choices to navigate communities and interpreted them as signs of investment. In contrast to algorithmically curated platforms, Discord requires users to actively seek out and join communities through direct invites or intentional searches, and participants appreciated this for fostering more purposeful interactions. P19 explained, \inlinequote{I get to choose the Discord servers I'm in and the people that are in there. So it makes it easier [to build connections] because I know I'm choosing this group of people\ldots{} to at least try and start interactions and relationships with.} P09 contrasted this with Instagram's algorithm-driven approach, which they compared to a \inlinequote{quilt} of mainstream topics where interactions follow patterns laid out for the user. On Discord, users can \inlinequote{pick which string} or \inlinequote{which pattern} they want to follow, leading to more focused engagement.

Clear structure was especially important in larger servers. P19 compared joining one to \inlinequote{walk[ing] into a room with a thousand people I didn't know\ldots{} and just having to figure out how to introduce myself to people.} Organized channels and roles made these spaces easier to navigate and signaled investment in the community. Disorganization sent the opposite signal. P19 explained: \inlinequote{I've seen servers where the only roles are admin and non admin\ldots{} and those servers typically only have a few text channels like General\ldots{} and for me, that shows a lack of structure. It's absolute chaos. I can't track who's who. I can't track the conversation. I don't think I've been in one of those servers longer than 12 hours before I've left.}

Community control also let members reshape or exit their environment. P09 noted that on Instagram, \inlinequote{the loudest people tend to determine what happens} and users \inlinequote{cannot change that environment}. On Discord, communities can \inlinequote{ban} toxic individuals, and users can \inlinequote{just join a different server with the same topic.} These powers supported a sense of ownership and helped communities establish local norms. They also created hierarchies that could undermine social equality: roles order the member list, moderators hold sanctioning authority, and permissions determine who can see and say what (\S\ref{sec:background}). The same powers that let members shape a community can also exclude people from it.

\subsection{Making Participation Controllable, Reversible, and Low-Obligation}
\label{sec:controllable-participation}
Alongside returnable settings, participants described controlling how they entered interactions, presented themselves, shifted modalities, and withdrew. Shared interests eased initiation, flexible identity options reduced performance pressure, graduated entry let participants evaluate communities and people before committing, and playful norms and broad access kept participation low-stakes. Each arrangement supported some third-place functions while creating tensions with others.

\subsubsection{\texorpdfstring{Shared Interests Made It Easier to Start Conversations}{Shared Interests Made It Easier to Start Conversations}}
\label{sec:shared-interests}
Shared interests and activities gave strangers natural reasons to speak. Joining a themed server already communicated common ground, while recurring activities created easy openings. P06 described one interactive game as \inlinequote{a really fun game to get to know other people.} Servers often featured \inlinequote{Question of the Day} channels, where bots posted prompts such as \inlinequote{`What's your favorite ice cream flavor?'} P25 said, \inlinequote{People are gonna talk and be like, `Oh, I like that flavor too.'} P03 described a \inlinequote{counting bot} as \inlinequote{the most active channel on the server}; members kept the simple activity interesting by using math, so \inlinequote{Instead of saying 19, they might say 20 minus 1, and it still counts.}

Thematic organization around games, academic subjects, hobbies, or shared identities also made it easier to approach strangers. A user's presence in a server or channel signaled \inlinequote{some kind of mutual interest\ldots{} off the bat} (P19), helping participants assess compatibility. P23 noted the contrast with direct messaging: \inlinequote{in messages, I would never know if someone likes music unless I ask them directly.} This common ground offered \inlinequote{some level of similarity to get along with} (P24), especially in servers devoted to \inlinequote{very niche hobbies} (P22). P11 compared Discord to \inlinequote{the wallflower at the party\ldots{} talking to somebody else who's also a wallflower\ldots{} And you're like, `Hey, this party kind of sucks.' `Yeah, it does kind of suck.' And you kind of bond off of that shared resentment of the party.}

Shared interests also sustained conversation. P08 said themed channels \inlinequote{giv[e] people things to talk about\ldots{} just kind of like ideas or conversation starters.} P21 appreciated finding \inlinequote{people who shared my interests} because \inlinequote{we always have something to talk about.} P12 explained why approaching someone without such common ground could feel awkward: \inlinequote{You don't usually just stumble across someone and say, `Hey, do you want to be my friend?'\ldots{} I think you get a couple of weird looks if you [do that].} Repeated encounters then made others familiar. P12 compared this process to school: \inlinequote{you see the same people around campus, and you get familiar with their faces, kind of the same way that you would be familiar with the profile picture of someone you keep seeing in your chat rooms.}

Conversations could also expand beyond their original topic. P02 compared Discord to choir socials: \inlinequote{we don't have to be talking about singing, but we're just united through that\ldots{} and we just have a good time.} Shared interests supported conversation, play, and the formation of regulars, but they could also encourage conformity and insularity. P03 noted that the same community mechanisms that \inlinequote{uplift people} and \inlinequote{provide sense of community} \inlinequote{can also\ldots{} create friendships in which\ldots{} the goal is to\ldots{} radicalize} others.

\subsubsection{\texorpdfstring{Identity Flexibility Supported Comfortable Self-Expression}{Identity Flexibility Supported Comfortable Self-Expression}}
\label{sec:identity-flexibility}
Contextual identity let participants vary how identifiable they were in each community. They generally used stable pseudonyms, adjusting how much offline information they disclosed according to the setting. Reduced visual identifiability mattered most for appearance-based pressure. P13 shared, \inlinequote{I cared way less anonymously on any Discord server\ldots{} But I can't in person. When I try making friends, I'm consistently thinking someone's eyes are on me.} P13 felt that text shifted attention toward conversation: \inlinequote{you tend to not judge a book by its cover because you don't know who they are\ldots{} So you tend to make friends based on text or just personal interactions.} For P13, this also made compatibility easier to assess: \inlinequote{When I started college, everyone seemed super friendly but also super fake. But online, because they were anonymous, they were their true selves. And that's kind of easy for you to tell right off the bat whether you wanna hang out with this person\ldots{} or not.} Reduced identifiability could also permit misrepresentation (\S\ref{sec:controllable-presence-discussion}).

Discord's profile system also let users shape their presentation across communities. P25 noted, \inlinequote{My profile is different\ldots{} in every server\ldots{} I have so many different profiles. Those are all me\ldots{} I think it gives you a lot more personality, and it does stimulate conversation.} Participants used this flexibility for practical and playful purposes. P05 could \inlinequote{attach my real name} in school-related servers \inlinequote{so that way people would know who I am}, while P15 described a profile where \inlinequote{one says, `student academic weapon' and `academic victim' depending on the hour of the day.} P10 observed, \inlinequote{There's no link to who you actually are\ldots{} Discord kind of curates a space where you don't have to do that.} P04 valued the freedom to explore identity: \inlinequote{I can be anyone I want without other people judging me or knowing who I am.}

This flexibility reduced the prominence of cues such as appearance, wealth, and offline standing, but communities produced visible local status of their own. Roles order the member list, moderators carry sanctioning power, Nitro badges signal payment, and tenure confers standing (\S\ref{sec:background}). As P02 explained, \inlinequote{Usually, the moderators are people that are veterans, like long-time members\ldots{} and are also really involved in the community.} Contextual identity therefore supported the leveler and low-profile functions while also creating new hierarchies and risks of weak accountability (\S\ref{sec:design-principles} and \S\ref{sec:controllable-presence-discussion}).

\subsubsection{\texorpdfstring{Graduated Entry Let Participants Assess Compatibility Before Committing}{Graduated Entry Let Participants Assess Compatibility Before Committing}}
\label{sec:graduated-entry}
Participants could regulate how they entered a community: previewing a server before joining, observing after joining, contributing when ready, and moving between text and voice as trust grew. They could also remain peripheral, step back, or leave. We use \textit{graduated visibility} to describe this control over how observable one's participation becomes to others (\S\ref{sec:controllable-presence-discussion}).

P08 contrasted this control with platforms where approaching others feels \inlinequote{less in control} and \inlinequote{like a stranger.} On Discord, users were \inlinequote{already within the group} and could \inlinequote{choose to join after observing}, making it feel as though \inlinequote{you kind of know them a bit.} P13 valued a \inlinequote{trial period in the server} where they could \inlinequote{talk to people\ldots{} without committing to a friendship}; as they put it, \inlinequote{even if it's anonymous, I kind of know this person's not a murderer.} Others valued being able to \inlinequote{lurk} (P12, P19), \inlinequote{get the vibe} (P12), and present themselves more confidently when ready. Graduated entry supported low-obligation participation and accessibility. It could also weaken mutual accountability, and low-cost exit meant connections could dissolve without notice or repair (\S\ref{sec:synthesis}).

\subsubsection{\texorpdfstring{Conversational, Playful Norms Reduced Pressure to Perform}{Conversational, Playful Norms Reduced Pressure to Perform}}
\label{sec:playful-norms}
Participants described chat-centered interaction as conversational and low-stakes. P25 described Reddit as \inlinequote{someone asking a question or sharing a picture, and people responding to the OP, but they're not really talking to each other.} She experienced TikTok as \inlinequote{restricting communication-wise}, Facebook groups as requiring users to \inlinequote{make an entire post}, and Twitch as offering \inlinequote{just live streaming} and \inlinequote{quick chat} where, in her view, \inlinequote{you can't form relationships.} P04 said these differences \inlinequote{in the short term, might not make a lot of difference}, but over time Discord made users feel \inlinequote{a part of a community.}

Voice channels supported faster exchange. P12 valued their \inlinequote{instant response} because it was \inlinequote{easier to tell} what others were like. The same infrastructure supported larger events, including \inlinequote{talent shows} where 200--300 participants could be \inlinequote{singing, show[ing] talents, and talk[ing]\ldots{} at once} (P03), which \inlinequote{really builds a sense of community bonding} and \inlinequote{develops a sense of cohesiveness.}

Servers and message histories persisted, but individual messages carried little weight within an ongoing conversation. P19 explained: \inlinequote{It's a lot\ldots{} more casual on Discord\ldots{} On Facebook groups, if I'm gonna make a post, I have to have something to say\ldots{} I am contributing something, and then everyone's responding to me\ldots{} But on Discord\ldots{} I'm saying my thing, but then so is the next person, and so is the next person\ldots{} It is back and forth and doesn't feel centered on me.} P04 similarly noted, \inlinequote{you see all those four or five different people constantly chatting. And you are also part of that group.}

Participants also connected Discord's gaming origins to its playful atmosphere. P19 observed, \inlinequote{Discord's big thing is being a big community for gamers.} P18 described it as \inlinequote{kind of its own thing, where it's kind of just to have fun conversations, and like it's not as serious.} GIFs, memes, and custom emotes helped users express emotion and build local culture. P09 described how students \inlinequote{like to send GIFs in after an exam. It's like `Oh, I'm in danger'\ldots{} it's just a way to\ldots{} bond over how hard the class is, and\ldots{} lowers the tension.} P03 noted, \inlinequote{in our server specifically\ldots{} we have a Duolingo bird which we name `Fear'\ldots{} different servers have different things.} These practices supported play, conversation, and a home away from home. They could also make hostile exchanges easier to dismiss as jokes (\S\ref{sec:design-principles}).

\subsubsection{\texorpdfstring{Open Access and Low Barriers Made It Easier to Meet People}{Open Access and Low Barriers Made It Easier to Meet People}}
\label{sec:low-barriers}
Discord lowered geographic and social barriers to meeting people. Shared server contexts and username-based accounts often let users contact one another without a prior offline relationship or extensive personal information, subject to individual privacy settings and server restrictions. Participants described servers as \inlinequote{more out in the open} (P21), where they could \inlinequote{shoot message[s] directly without hitting a friend request} (P04). Communicating without exchanging phone numbers felt safer to P05 and made initiating friendships easier for P07, P10, and P21.

Participants contrasted this openness with other platforms. P04 described navigating \inlinequote{100 different posts}, having to \inlinequote{filter out} interests, sending a friend request, and then having to \inlinequote{wait for 2-3 days to connect} on Facebook. P25 valued interacting with a \inlinequote{big community} beyond an immediate circle, unlike Skype's friend-only model. P12 called the resulting connections \inlinequote{incidental friendships}, contrasting them with Instagram, where users \inlinequote{seek out\ldots{} individual people}. P11 compared Instagram to \inlinequote{a private party} where \inlinequote{you're trying to get in the party, and you can't get in the party. So you get sent to the back of the line.} On Discord, users were already \inlinequote{in the party, and then you're talking to somebody who also doesn't like the party. And it's like, `Hey, you know, let's just start talking.'}

Public, discoverable servers offered topic-based communities, while other servers remained invitation-only. P04 said, \inlinequote{I can find a server\ldots{} for whatever purpose\ldots{} there's something for everything, something for everyone.} P12 similarly believed that \inlinequote{with enough time and patience, you can almost always find a community, regardless of the topic.} Access across operating systems and devices lowered technical barriers. P15 described Discord as \inlinequote{easily accessible} because users could \inlinequote{access it on multiple devices at once}, and P08 explained that many young users without phones relied on it. P07 contrasted this with iMessage's limitation to Apple's ecosystem.

Open voice channels lowered temporal and social barriers. P04 appreciated being able to \inlinequote{just drop in and drop out of a particular voice channel at any point without taking anybody else's consent.} P16 similarly noted, \inlinequote{You can leave without disturbing the call\ldots{} or if you can't join at the beginning,\ldots{} you can just join in later.} P07 summarized, \inlinequote{That's why I like Discord better than Skype---you don't have to request to join or ask someone to add you to the call. You just click on the [channel]\ldots{} and join.} This ease of entry and exit supported spontaneous interaction and accessibility. It also made young people easier for strangers to reach, creating safety concerns discussed in \S\ref{sec:controllable-presence-discussion}.

\subsubsection{\texorpdfstring{Keeping Discord Open in the Background Supported Companionship}{Keeping Discord Open in the Background Supported Companionship}}
\label{sec:background-use}
Several participants kept Discord open throughout the day and said this background presence provided companionship. P07 said, \inlinequote{I always have it on in the background\ldots{} even though I'm not using it, it's just always there,} keeping it on a second monitor and briefly joining voice channels when he saw friends playing. P02 described always-open voice lobbies as rooms one could inhabit: \inlinequote{it truly is like a room almost\ldots{} you could just live there\ldots{} and if someone joins in you just greet them.} They added, \inlinequote{You can just stay in that call forever. It constantly is just open.}

P08 connected this norm to low-pressure relationship maintenance. Because \inlinequote{everyone has\ldots{} discord on in their background}, a slow reply carried little meaning: \inlinequote{It's just right there, like you respond whenever.} P04 valued being able to \inlinequote{plug in discord in the background} on his game console during sessions with friends. Participants managed their attention through granular controls. P02 noted, \inlinequote{You don't have to mute the whole server. You could just mute channels or specific things\ldots{} or people.} P20 experienced these controls differently. Muting and deafening made mediated presence feel thinner because \inlinequote{in real life you don't have that option to turn off your video, to deafen for 5 min.}

\subsection{How the Two Conditions Worked Together as Relationships Deepened or Dissolved}
\label{sec:synthesis}
The arrangements in \S\ref{sec:returnable-settings} and \S\ref{sec:controllable-participation} worked together over time. Three participants' accounts show how returnable settings and controllable participation accompanied deepening relationships. Other accounts show where this process stalled or reversed: encounters without repeated contact expired, and persistent settings that felt obligatory did not support comfortable participation.

P19 gave the most detailed account of this progression. They joined a comedian's 2{,}000-member server to support a streamer friend, then limited their exposure: \inlinequote{I really just stayed to my 2 or 3 text channels for the longest time,} and \inlinequote{I lurked for a while, so I didn't really say too much when I very first started.} The persistent setting allowed familiarity to accumulate while they remained peripheral. Eventually, they \inlinequote{would vaguely have an idea of like, okay, when these 3 or 4 or 5 people are active. I can crack a joke\ldots{} and they'll probably at least be kind of interested about what the newbie has to say.}

P19 then became an active contributor: \inlinequote{I ended up contributing like over 200 quotes to that channel.} Contact also became more personal: \inlinequote{we went from talking, maybe\ldots{} saying one thing to each other in the chat to talking a lot at DMs, which was how we became really close friends.} P19 called these \inlinequote{some of the best friendships I have for the longest time} and said of one friend who lived nearby, \inlinequote{Once we met up in person we were inseparable for like 3 years.} At each stage, the persistent setting created another opportunity to interact, while P19 decided whether and how to participate.

P23's account began with grief. After her partner died, she was invited into a private group chat with his friends. She initially remained at the periphery: \inlinequote{I didn't talk on it very much. Just because I didn't feel comfortable with it.} Because the group remained available, their exchanges gradually increased over several years until, in her words, \inlinequote{We start texting like a lot.}

P24's account shows how this process could lead to community belonging. She entered a disability-related server through its rules-and-verification gate and mentioned in the general chat that she \inlinequote{couldn't do my geometry Delta math}. Another member helped, and \inlinequote{we talked back and forth on the server for a little while.} Their contact moved to Instagram and then to text, settling into weekly check-ins with someone \inlinequote{fairly close\ldots{} to me.} The community's inner circle later built the house-structured server she described as \inlinequote{a group of us that are more close knit}. Her \inlinequote{bedroom} channel holds pictures of her cat and \inlinequote{quotes of weird things my family has said}, expressing belonging through a personal room within a shared setting.

The process could also run backward. P18's closest online friendship formed in a fandom group chat and ended after they created a Discord server for voice chat: \inlinequote{we also made a discord server\ldots{} so we could like voice chat,} and \inlinequote{I think once we did that it kind of fell off because we would use discord more, and then we just kind of stopped talking.} For P18, moving from text to voice coincided with the relationship ending, although other participants experienced that move as a sign of deepening. Encounters without a shared setting simply expired. P04 said of a contact from a plane-spotting exchange, \inlinequote{I won't even be in contact with that person after this,} while P20 left servers within half an hour and described his Discord episodes as having died out. P06 summarized the pattern: public-server connections \inlinequote{die out really quickly} unless something more personal follows.

Most encounters participants described never deepened. P07, for example, said of many contacts, \inlinequote{I'm not really close with these people anymore, only the server.} Low-obligation participation made it easy to let a connection lapse, so relationships deepened only when both people continued to choose them. Persistent settings kept that choice open across weeks and years.

Some relationships followed different paths. P03 mostly deepened friendships that predated Discord. P22, whose earliest ties formed online (\S\ref{sec:close-relationships}), now meets people in person before adding them on Discord. The accounts above therefore describe one recurring pattern in participants' histories, not every route their relationships took.

%% file: sections/5_discussion.tex
\section{Discussion}
\label{sec:discussion}

Our findings suggest one possible mechanism: participants described relationships forming through persistent, socially legible settings combined with controllable, reversible participation. We explain how this mechanism operated across four levels of design, relate it to Oldenburg's characteristics, and examine the opportunities and risks of controllable presence.

\subsection{\texorpdfstring{How Returnable Settings and Controllable Presence Work Together}{How Returnable Settings and Controllable Presence Work Together}}
\label{sec:returnability-discussion}
Communities become returnable places through stable structures and shared norms, while users control how they participate. This echoes Alexander's intimacy gradient, which orders spaces from public to private to support social life~\cite{Alexander1977-pattern}. Participants described a similar range in channels that extended from open lobbies to small, trusted rooms.

Participants' accounts distinguish four levels. The platform supplies basic features and policies: servers, channels, roles, permissions, presence indicators, and moderation tools. Server operators configure them into bounded spaces with particular structures, hierarchies, and entry rules. Communities turn these spaces into places through norms, rituals, shared history, and local reputations. Individuals regulate their visibility, identifiability, and commitment within and across places. P19's account of an almost unstructured server (\S\ref{sec:community-control}) shows how the same platform features can succeed or fail depending on how a server is configured and inhabited. Platform-level guidance alone misses this difference.

The two sets of conditions matter together because each supplies what the other lacks. Persistence without participation control turns continuity into exposure or obligation; control without continuity leaves encounters with nothing to return to. Both failures appeared in our data (\S\ref{sec:synthesis}): returning became a chore for P01, while P04's and P20's low-stakes encounters expired without repeated contact. Together, these conditions let familiarity accumulate through recurring, low-obligation encounters and allow people to deepen contact after assessing compatibility. Most encounters remained shallow precisely because reversible participation also made them easy to leave.

Participants also explained the difference by comparing Discord with feed- and profile-centered platforms. There, they said, a comment thread scrolls away and the next encounter occurs elsewhere, before a different audience (P12, P19; \S\ref{sec:returnable-settings} and \S\ref{sec:controllable-participation}). Their spatial analogies---town squares, cafes, and a house with rooms---expressed returnability, while lower-stakes initiation reflected participation control. These comparisons reflect participants' perceptions of the platforms they used.

Discord helps us examine the role of embodiment because it lacks avatar bodies, a navigable world, and required synchronous interaction, yet participants still described third-place experiences. It separates returnability and participation control from visual realism while offering voice, video, streaming, roles, and paid status signals. Both embodied worlds and communication platforms can host third places; here, relationship formation centered on returnability and participation control.

\subsection{Design Principles for Virtual Third Places}
\label{sec:design-principles}
Oldenburg's concept of third places identified eight characteristics that distinguish effective informal public gathering spots (third places) from homes (first places) and workplaces (second places)~\cite{Oldenburg1999-nd}. Prior work on MUDs, chatrooms, and online forums established decades ago that text-based environments can foster place-like experiences and support relationship formation~\cite{Baym1995-ue, Bruckman1995-ge, Rheingold1993-vc}. Building on this work, we use Oldenburg's characteristics as social functions and derive design principles for how platforms and communities can support them. Table~\ref{tab:oldenburg} summarizes each principle and the functions and tensions associated with it.

Each design principle can support several functions, sometimes while undermining another, and each function usually depends on several principles. Two characteristics illustrate this pattern. Pseudonymity can support the \textit{leveler} function by reducing cues about appearance, wealth, and offline status. Discord communities also create local hierarchies through roles, moderator authority, paid Nitro signals, and tenure (\S\ref{sec:background} and \S\ref{sec:identity-flexibility}), so reducing offline status cues does not eliminate status. A \textit{home away from home} emerges from persistence, intimate subspaces, customization, and governance together. The governance that makes insiders feel at home can also exclude newcomers (\S\ref{sec:community-control}). Each row of Table~\ref{tab:oldenburg} therefore shows the functions a design principle can support and the tensions it can create.

\input{inserts/oldenburg}

The design principles in Table~\ref{tab:oldenburg} can apply across different media. A feed-based platform adding community spaces and an embodied world designing gathering places may face similar trade-offs. Their relevance will vary because content-centered platforms such as Instagram and TikTok serve different purposes and prioritize different outcomes~\cite{zhang2024form}.

Participants formed and deepened ties without embodied avatars, navigable space, or simulated bodies. Limited embodiment may even help support three of Oldenburg's functions. Limited self-presentation can support a \textit{low profile}; making appearance and dress less visible can reduce imported status cues and support the \textit{leveler}; and modest device and skill requirements can improve \textit{accessibility and accommodation} (\S\ref{sec:low-barriers}). Comparative studies are needed to test these possibilities. Embodiment also provides nonverbal expression, spatial coordination, and a strong sense of co-presence~\cite{Smith2018-communication}. Discord itself combines text with voice, video, streaming, roles, and paid status signals, so our case does not isolate text. Long-standing 3D worlds host communities with similar social arrangements~\cite{Boellstorff2008-coming}, and text channels and embodied spaces can complement each other~\cite{Alcala2023-zy}. Platforms such as Second Life may be limited as much by hardware demands, software complexity, and learning curves as by their design. As generative tools and maturing XR hardware make greater realism a common response to social design problems, our findings suggest that designers should first consider whether people have a place to return to and enough control over how they participate.

\subsection{Controllable Presence: Graduated Visibility, Contextual Identity, and Reversible Participation}
\label{sec:controllable-presence-discussion}
By \textit{controllable presence} we mean participants' ability to regulate whether and how they become socially available, to whom, and with how much obligation. The accounts in \S\ref{sec:controllable-participation} show six reversible dimensions of this control: observability (reading without posting, reacting, posting, entering active exchange), identifiability (pseudonymous handle, server-specific nickname, disclosed offline identity), audience (public channel, role-restricted channel, small group, direct message), modality (text, voice, video, in-person contact), attention (backgrounded or focused participation, muting and deafening), and commitment (previewing, joining, recurring participation, leaving, returning). We present these dimensions as a qualitative description; future work could test them as a scale. Participants did not have to move through them in order and often remained at one level. Hollan and Stornetta argued that mediated communication should stop imitating physical co-presence and instead support what co-presence cannot~\cite{Hollan1992-px}, and Soukup made the parallel argument for virtual third places~\cite{Soukup2006-sb}. Controllable presence extends that argument to third-place design. Digital settings can separate dimensions of presence that physical settings bundle together, and participants treated this flexibility as central to forming relationships.

We use \textit{graduated visibility} for the observability dimension alone: control over how perceptible one's participation becomes to others across a continuum from reading without posting to visible contribution and active exchange. Graduated visibility differs from four neighboring concepts. Social translucence concerns how systems make \textit{others'} activity visible to support coordination~\cite{Erickson2000-gn}; graduated visibility concerns control over one's \textit{own} exposure. Barta and Andalibi's visibility objects describe self-presentation choices on profile-centered platforms~\cite{Barta2024-visibility}; the control we describe operates within persistent communities. Lurking and legitimate peripheral participation describe how people develop \textit{competence} and community membership~\cite{Lave1991-situated, Nonnecke2000-lurker}; our participants also remained peripheral to manage \textit{disclosure and exposure}. Context collapse occurs when different audiences meet under one identity~\cite{Marwick2011-lr, Vitak2012-ne}; graduated visibility helped participants keep those audiences apart. Prior research shows that users create this control through finstas, alternate accounts, throwaway accounts, and platform migration~\cite{Tao2023-an, Ellison2016-yp, Haimson2016-zr, Stutzman2012-ev}. Discord builds in much of it: per-community nicknames and reversible exposure require no second account and break no rules. However, per-server avatars and bios require a paid subscription, and the platform-wide account and mutual-server lists remain visible beneath each local presentation.

Minors and emerging adults alike described identity flexibility as useful, although the study was not designed to compare age groups. Adolescence and emerging adulthood are important periods for identity formation, during which individuals experiment with different self-presentations, seek feedback from peers, and work toward a coherent sense of self~\cite{erikson_identity_1994, Marcia1966-development}. Experimenting with identity can bring the risk of rejection, and peer feedback during this period weighs heavily on self-concept~\cite{Harter1990-causes, Steinberg2005-qy}. Physical spaces constrain such exploration because bodies communicate social information continuously and involuntarily---appearance, voice, clothing, and comportment shape others' perceptions before any intentional self-presentation occurs. For adolescents experiencing heightened self-consciousness about their changing bodies, this embodiment raises the stakes of every social encounter~\cite{Elkind1967-dw, Sebastian2008-development}.

In participants' accounts, controllable presence eased these developmental pressures in three ways. Making appearance less visible helped participants connect through shared interests and conversation during initial interactions. Pseudonymous usernames separated online identity from offline reputation, reducing anxiety about how experimental self-presentations might affect established relationships or future opportunities. Server-specific profiles let users emphasize different aspects of themselves across communities---academic seriousness in a study group, creative enthusiasm in an art server, playful irreverence among gaming friends---without forcing these presentations together. Identity research likewise treats contextual variation in self-presentation as ordinary~\cite{Turkle1995-life, Marwick2011-lr}. Because users define authenticity on social media in conflicting ways~\cite{Haimson2021-qd, Barta2021-yh, Li2025-befake, kim2024bereal}, we interpret these accounts as comfort and fit within a particular context.

The developmental value of this control becomes clearer on platforms that enforce singular, persistent, real-name identities. On such platforms, every interaction contributes to a cumulative public record visible to diverse audiences---peers, parents, future employers---creating what Marwick and boyd term ``context collapse''~\cite{Marwick2011-lr}. Youth must then self-present simultaneously to audiences with conflicting expectations, often defaulting to cautious self-censorship that narrows what they share~\cite{Vitak2012-ne}. Discord's community boundaries and server-specific presentation can mitigate some forms of context collapse. Users can explore different aspects of identity in different communities~\cite{erikson_identity_1994, arnett2000emerging}. Mutual-server lists, platform-wide accounts, screenshots, durable message history, and ties that span contexts can still reconnect audiences, so this separation remains partial.

Beyond enabling identity exploration, Discord's geographic reach reduces a constraint of physical third places: dependence on nearby people. Many adolescents lack independent transportation and depend on adults to reach spaces outside their neighborhoods~\cite{Skelton2000-nothing}, and their sense of connection depends partly on what their local communities offer~\cite{Marcu2026-adolescent}. Their potential friends may be limited to people at the same school or in the same area, and they may find no local peers who share particular interests or identities. Discord lets youth find such communities regardless of location. This matters especially for youth with stigmatized identities or those exploring aspects of themselves privately~\cite{Craig2014-you, McConnell2018-identity, Fox2016-queer}. The same reach also makes minors more accessible to strangers, including adults; we address this risk below.

Together, graduated visibility, contextual self-presentation, and geographic reach created conditions participants described as comfortable and low-pressure for forming ties. When connections form through sustained interaction around shared interests rather than physical proximity or surface-level characteristics, and when identity experimentation feels less risky, youth can engage more fully. Community status systems had mixed effects. Recognition came from participation rather than imported social advantages, but roles, moderator authority, and tenure created a new local hierarchy~\cite{Yoon2025Communitieso}. These structures could also offer developmental value by giving younger users opportunities to practice leadership~\cite{Yoon2025Communitieso}.

Every property described in this subsection can be used for benefit or harm. The steps that let a cautious newcomer enter gradually can also enable grooming: quiet observation, pseudonymous rapport, movement into private channels, shifts in modality, and requests for personal information. Recent work documents both possibilities on Discord: teens use vigilance, safety tools, and selective participation to manage risky interactions~\cite{Koung2026-teen}, and they develop privacy norms collectively within communities~\cite{Dong2026-collective}. Safety design should therefore protect users at each step without eliminating their control. Possible measures include age-based defaults for who may initiate contact, role- or tenure-gated direct messages, consent before changing modalities, added friction or verification at higher-risk transitions, optional presence indicators, slow modes, silent blocking and leaving, transparent governance with appeals, and moderation records that identify repeated harm~\cite{Wisniewski2018-rc, razi2020privacy, Agha2023-mu, Schoenebeck2021-oh}. These proposals come from prior research; our study tested none of them and cannot show how often predators use this sequence.

\subsection{Limitations and Future Work}
\label{sec:limitations}

All participants had reported forming at least one friendship on Discord, so the sample centers people for whom the platform supported connection. The data cannot establish how often Discord use leads to friendship, whether Discord supports friendship more effectively than other platforms, or how people experience failed or harmful connections. Harms are underrepresented: participants mentioned risk, surveillance, and exclusion in passing (\S\ref{sec:returnable-settings}--\S\ref{sec:synthesis}), and Table~\ref{tab:oldenburg}'s harm column draws from prior research rather than evidence from this study~\cite{Jargon2019-fq, Kelly2022-nl, Koung2026-teen}. Because the same design principles can produce benefits and harms, studying the trade-off requires sampling for both. Future studies should recruit people who left Discord, failed to connect there, or were harmed there.

The data are retrospective self-reports. Participants reconstructed relationship histories, and their platform comparisons reflect the interpretations of people for whom Discord worked. We cannot distinguish arrangements that caused their connections from those that this self-selected group happens to enjoy. The trajectories in \S\ref{sec:synthesis} follow sequences within individual accounts, but those sequences were remembered rather than observed. Longitudinal studies following newcomers into servers, and comparative studies following similar users across platforms, could test whether the proposed pathway predicts relationship formation over time.

The settings and sample varied in ways our study could not compare directly. Participants' servers ranged from ten-member friend groups to communities of thousands, with governance ranging from a single administrator to layered role systems. Our findings span this variation, so future work should compare which forms of governance produce particular benefits and problems. The 15--24 age range also spans distinct developmental periods. Without younger, older, or adult comparison groups, we cannot claim that these findings are unique to youth. Recruitment through one university's contact lists and campus, with first-come scheduling, further limits who is represented. The sample is also demographically concentrated: 68\% selected Girl/Woman and 56\% identified as Asian or Asian-American, so these groups are over-represented.

Finally, our interview questions and analytic framework shaped the data. The protocol centered on Discord and on friendships participants had already formed, one question invited platform metaphors (\S\ref{sec:procedure}), and the analysis applied Oldenburg's characteristics deductively. We retained material that did not fit the framework (\S\ref{sec:data-analysis}), but a study without a third-place lens, or one that samples across platforms, could show how much our findings reflect these choices. Future studies could test our claims by sampling unsuccessful and harmful cases, following relationships over time, comparing platforms, governance models, and age groups, or analyzing the same topic without a third-place framework.

%% file: inserts/oldenburg.tex
{
\small
\begin{longtable}{>{\raggedright\arraybackslash}p{2.6cm} >{\raggedright\arraybackslash}p{4cm} >{\raggedright\arraybackslash}p{2.9cm} >{\raggedright\arraybackslash}p{3.1cm}}
\caption{Design principles for virtual third places. Each principle may support or strain several of Oldenburg's third-place functions, and each function may depend on several principles. The evidence column points to the Results and Background. Potential harms are drawn from prior research.}
\label{tab:oldenburg} \\
\toprule
\textbf{Design principle} & \textbf{Relevant evidence} & \textbf{Third-place functions} & \textbf{Potential trade-offs or harms} \\
\midrule
\endfirsthead

\toprule
\textbf{Design principle} & \textbf{Relevant evidence} & \textbf{Third-place functions} & \textbf{Potential trade-offs or harms} \\
\midrule
\endhead

\bottomrule
\endfoot

Keep community spaces and shared histories available over time & Persistent servers and channels let participants return to the same setting, conversations, and people. Participants associated repeated encounters with growing recognition and relationships (\S\ref{sec:persistent-spaces}) & Supports: regulars; home away from home; conversation. Can strain: neutral ground, because shared history can create standing and obligation & Durable records and repeated access can enable surveillance, sustained conflict, and grooming~\cite{Jargon2019-fq, Koung2026-teen} \\
\addlinespace

Segment interaction while keeping related spaces connected & Channels and categories organize topics and group size ``under one roof.'' Smaller channels made repeated encounters more likely (\S\ref{sec:channel-segmentation}) & Supports: conversation; accessibility; home away from home. Can strain: neutral ground, because subspaces can develop insider groups & Fragmentation can hide harmful subspaces, divide communities, and overwhelm newcomers~\cite{Kelly2022-nl} \\
\addlinespace

Show activity and regulars without requiring interaction & Status, member-list, and voice-channel cues let participants gauge who was present and available before engaging (\S\ref{sec:visibility-cues} and \S\ref{sec:background}) & Supports: regulars; conversation; accessibility. Can strain: low profile, because visible activity can create performance pressure & Tracking, stalking, and pressure to appear available~\cite{Koung2026-teen, Dong2026-collective} \\
\addlinespace

Let communities shape their spaces through accountable governance & Roles, permissions, norms, and bots let communities shape culture and boundaries. Visible structure signaled investment, while its absence prompted participants to leave (\S\ref{sec:community-control} and \S\ref{sec:background}) & Supports: home away from home; regulars; playful mood. Can strain: leveler, because roles create hierarchy; neutral ground, through gatekeeping & Concentrated moderator power, gatekeeping, exclusion, and unclear appeal processes~\cite{Schoenebeck2021-oh} \\
\addlinespace

Organize interaction around recurring shared interests and activities & Topic-based servers and recurring prompts, such as question-of-the-day channels and counting bots, supplied common ground for starting and sustaining conversations (\S\ref{sec:shared-interests}) & Supports: conversation; playful mood; regulars & Conformity pressure, in-group hostility, and reinforcement of harmful views in tightly focused groups~\cite{Kelly2022-nl} \\
\addlinespace

Let users limit disclosure of appearance and offline status & Pseudonyms and server-specific profiles let participants connect through shared interests before disclosing offline identities. Participants adjusted how identifiable they were in each community (\S\ref{sec:identity-flexibility}) & Supports: leveler; low profile; accessibility. Can strain: leveler, because communities create new local hierarchies & Misrepresentation, catfishing, and weak accountability; local status systems can replace offline ones~\cite{Jargon2019-fq, razi2020privacy} \\
\addlinespace

Let people adjust participation and communication modes & Previewing servers, lurking, moving from text to voice, muting, joining and leaving calls, and leaving servers let participants regulate exposure and commitment while assessing others (\S\ref{sec:graduated-entry} and \S\ref{sec:background}) & Supports: neutral ground; accessibility; conversation; regulars & Covert observation, one-sided disclosure, grooming through gradual escalation, and ties that dissolve without repair~\cite{Jargon2019-fq, Wisniewski2018-rc} \\
\addlinespace

Support informal expression and shared play & Reactions, emotes, GIFs, humor, and recurring bot activities lowered performance pressure and built local culture (\S\ref{sec:playful-norms}) & Supports: playful mood; conversation; home away from home. Can strain: conversation, when noise crowds out discussion & Humor can be used to excuse harassment and discourage reporting~\cite{Schoenebeck2021-oh} \\
\addlinespace

Lower geographic, temporal, device, and learning barriers & Cross-device, asynchronous, free access expanded which communities participants could reach and when they could participate. Young people without phones could still take part (\S\ref{sec:low-barriers}) & Supports: accessibility and accommodation; neutral ground & Low barriers can also increase unwanted adult--minor contact and abusive entry~\cite{Jargon2019-fq, Kelly2022-nl, razi2020privacy} \\
\addlinespace

Let users move between background and focused participation & Participants kept Discord open in the background and described this presence as companionship that made it easier to rejoin interaction (\S\ref{sec:background-use}) & Supports: accessibility and accommodation; regulars through continuity of presence. Can strain: home away from home, when presence becomes an expectation & Eroded boundaries, pressure to remain available, and monitoring of idle presence~\cite{Koung2026-teen} \\
\end{longtable}
}

%% file: sections/6_conclusion.tex
\section{Conclusion}
\label{sec:conclusion}
Across 25 interviews, young Discord users described two arrangements that supported relationship formation. Persistent, socially legible settings gave them somewhere to return to, where they could find the same people and build a shared history. Controllable, reversible participation let them enter, observe, become more involved, and withdraw without making any step permanent. Participants associated the combination of these arrangements with repeated low-obligation encounters, growing familiarity, and the friendships and belonging they reported. Reversibility also kept most encounters shallow: relationships deepened only when both people continued to choose them, and the persistence of the setting kept that choice open.

Within controllable presence, we identify graduated visibility: the ability to move from observing to increasingly visible participation within a community, without creating secondary accounts. We also propose design principles that show how specific choices can support several of Oldenburg's third-place functions while creating tensions with others. Participants reported these experiences without realistic embodiment; comparative work is needed to examine what different forms of embodiment contribute.

These findings have broader implications for youth social connection. As physical third places become harder to access, persistent digital communities can offer young people recurring, low-pressure opportunities to meet others, explore identity, and develop belonging. Platforms can support these opportunities by giving people places to return to and control over how they participate, with safety protections at each step from observation to active participation. The same properties can also enable surveillance, unwanted access, one-sided disclosure, and grooming. Effective support for connection requires both user control and accountable safeguards.

%% file: sections/7_appendix.tex
\clearpage
\appendix
\label{appendix}

\section{Demographics Data}
\label{appendix:demographics}
\input{inserts/demographics-individual}
\clearpage

\section{Interview Protocol}
\label{appendix:protocol}
\noindent\textit{The protocol below is reproduced verbatim as administered, including colloquial phrasing.}
\input{inserts/protocol}

\clearpage

\section{Extended Platform Description}
\label{appendix:platform}
This appendix expands the platform overview in \S\ref{sec:background} with feature descriptions and interface mock-ups. Features are described as of the interview period (July--September 2024); see the scope note in \S\ref{sec:background}.

\input{inserts/screenshots/screenshot-2}

\parHeading{Servers and Channels}
Discord's basic organizational unit is the \textit{server}, a persistent virtual space reached through an invitation or search. Servers contain text channels for asynchronous communication and voice channels for synchronous audio and optional video~\cite{discord_beginner}. Administrators can group channels under labeled categories (Figure~\ref{fig:screenshot-1}(a)). Users can join multiple servers and reorder or group their icons in the left sidebar (Figure~\ref{fig:screenshot-2}(b)).

Channels can be public to all members or restricted by permission, allowing both open community spaces and private areas for moderators or particular groups.

\parHeading{Roles and Access Control}
Discord administrators use role-based permissions to govern what members can do~\cite{discord_roles}. Roles are labels assigned manually or automatically, with permissions ranging from sending messages to managing channels or banning users. Higher roles can modify lower ones, and members receive the combined permissions of every role they hold.

Roles also shape what users see. They can set username colors and group members in the right sidebar (Figure~\ref{fig:screenshot-1}(d)), making status visible. Mentioning a role notifies everyone who holds it, and self-assignable roles let members join interest or notification groups (Figure~\ref{fig:screenshot-1}(c)). These features create local hierarchies: administrators and moderators hold visible authority to mute, kick, or ban users; operators assign roles at their discretion; and paid Nitro subscriptions add profile badges, animated avatars, and cross-server custom emoji~\cite{discord_nitro}. This structure informs our discussion of Oldenburg's leveler characteristic (\S\ref{sec:identity-flexibility} and \S\ref{sec:design-principles}).

\parHeading{Rules and Verification}
Community servers often set entry requirements before granting full access. Discord's ``Rules Screening'' requires new members to accept server rules before sending messages, adding reactions, or messaging other members~\cite{discord_rules_screening}. Until then, their visibility and interactions are restricted.

Many servers supplement or replace this feature with bot-managed verification (Figure~\ref{fig:screenshot-1}(b)). New members may see only a rules channel until they react with an emoji or run a bot command to receive an access role. This step can deter automated accounts and prompt members to read community guidelines.

\parHeading{Text-Based Communication}
Discord's primary interface centers on text-based chatrooms. The main screen displays the active channel's message history as a vertically scrolling feed~\cite{discord_beginner}. Messages support rich formatting, embedded links, and file attachments.

\begin{figure}[t]
    \centering
    \includegraphics[width=0.8\textwidth]{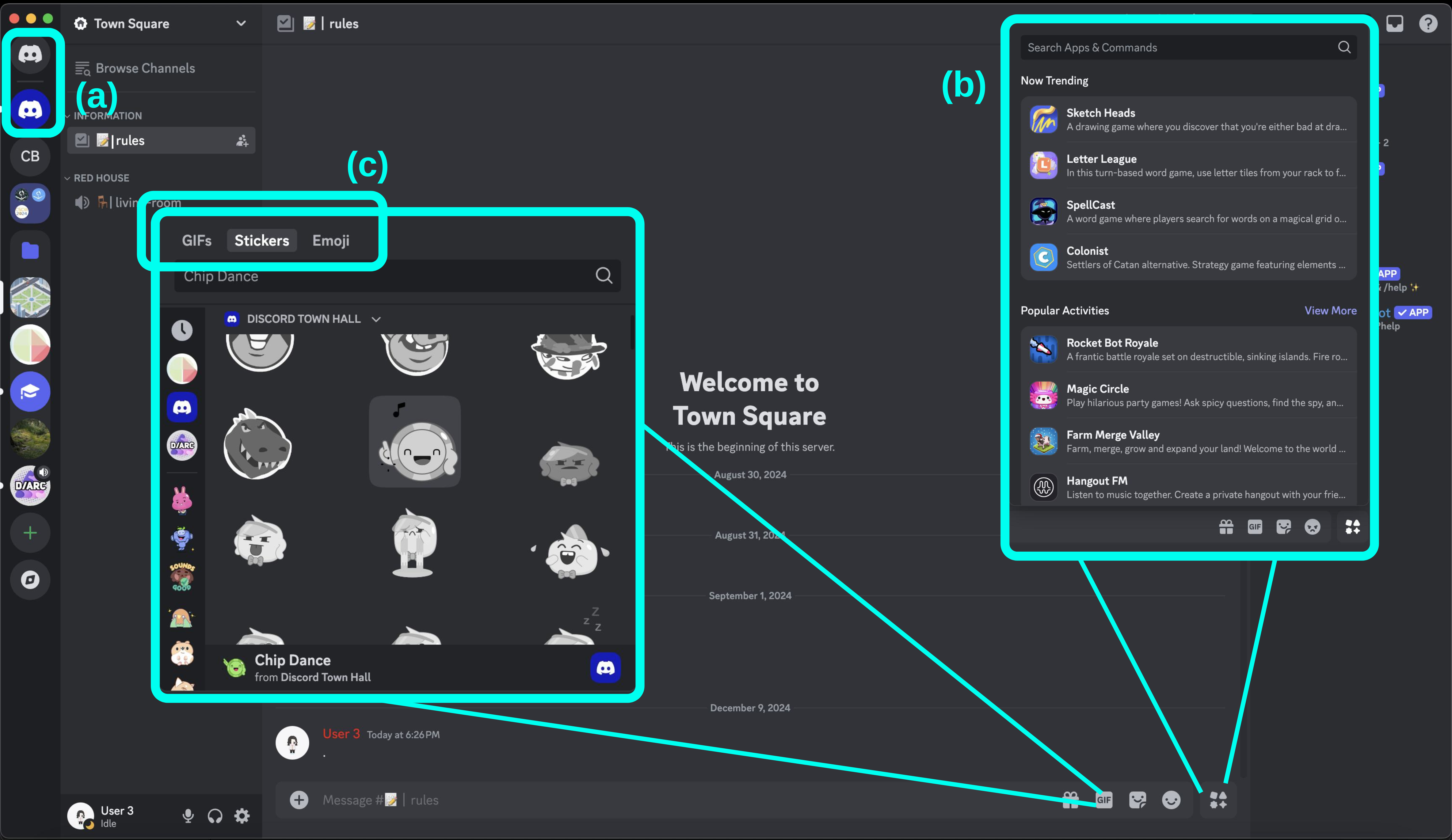}
    \caption{(a) Discord's logo; (b) Activities; (c) Message reactions and emotes.}
    \Description {This image shows a Discord interface with three highlighted sections labeled (a), (b), and (c). In section (a), the left sidebar is highlighted, showing a vertical list of server icons, with the Discord logo (that looks like a game controller) at the top and various server icons beneath it. In section (b), a panel titled ``Now Trending'' and ``Popular Activities'' is highlighted. It contains a search bar at the top labeled ``Search Apps \& Commands'' and displays a list of trending apps such as ``Sketch Heads,'' ``Letter League,'' and ``SpellCast,'' along with popular activities like ``Rocket Bot Royale,'' ``Magic Circle,'' ``Farm Merge Valley,'' and ``Hangout FM.'' In section (c), the sticker panel is highlighted, displaying tabs labeled ``GIFs,'' ``Stickers,'' and ``Emoji.'' The ``Stickers'' tab is selected, showing various stickers, including characters like a dancing figure and a dragon, with a search bar at the top and a label at the bottom that reads ``Chip Dance from Discord Town Hall.''}
    \label{fig:screenshot-reactions}
\end{figure}

Servers can upload their own custom emoji, which members use in messages and as reactions to others' messages~\cite{discord_emoji}. These server-specific emoji often become inside jokes and shared references within communities. Some communities develop recurring reaction patterns---greeting certain members with specific emoji combinations or reacting to announcements in prescribed ways---often coordinated through bots that can automate or enforce such behaviors (Figure~\ref{fig:screenshot-reactions}).

\parHeading{Voice Channels and Synchronous Interaction}
Voice channels operate on a ``drop-in, drop-out'' model that distinguishes Discord from traditional voice calling applications~\cite{discord_beginner}. Users click a voice channel to join immediately, with no call that rings recipients. The channel interface displays current participants, so members can see who is already present before joining (Figure~\ref{fig:screenshot-2}(d)). Other server members receive visual indication (and optional audio cues) when someone enters or exits a voice channel, facilitating awareness of ongoing conversations.

Voice channels support screen sharing, live streaming, and synchronized viewing, allowing members to broadcast their display or watch together (Figure~\ref{fig:screenshot-2}(a)); these capabilities reflect Discord's gaming origins while serving broader use cases like study groups or watch parties. Each voice channel also maintains an associated text chat (Figure~\ref{fig:screenshot-2}(c)), enabling participants to share links or continue discussion without interrupting voice conversation.

\parHeading{Activities and Bots}
Discord supports ``Activities''---built-in applications that allow users to play games, watch content, or engage in collaborative experiences directly within voice channels, without owning or installing separate software~\cite{discord_activities}.

Beyond built-in offerings, third-party bots extend what servers can do~\cite{discord_developer}. Bots are programmable automated accounts that respond to commands or events. Common applications include background music playback, automated moderation, custom interactive games, and integration with external services. For example, music bots join voice channels and stream audio audible to everyone present (Figure~\ref{fig:screenshot-2}(e)), creating collective listening sessions. Moderation bots filter problematic content, log events, or handle role distribution. Some communities use bots to support traditions such as automated greetings, scheduled discussion prompts, and gamified participation that become recurring parts of community culture.

\parHeading{User Profiles and Presence}
Clicking a user's name reveals their profile information (Figure~\ref{fig:screenshot-status}). The profile shows the user's chosen display name, avatar, and optional banner image. Users can write an ``About Me'' bio section and set a custom status message with an accompanying emoji. Discord also displays the user's current activity when available---for instance, the game they are currently playing and for how long, or the music they are listening to on a connected Spotify account.

\input{inserts/screenshots_2}

Users can also display their pronouns on their profile. Profiles additionally show contextual information about the viewer's relationship with the user: mutual servers (servers both users belong to) and mutual friends appear as lists. Each user has an online status indicator (online, idle, do not disturb, or invisible) visible to others, with controls for limiting this visibility. Within servers, users can have server-specific nicknames that differ from their global display name, and their profile displays which roles they hold in that particular community. Server-specific nicknames are available to all users; full per-server profiles, including a different avatar, banner, and bio per server, are a paid (Nitro) feature.

\parHeading{Server Discovery and Lurking}
Discord provides a ``Server Discovery'' feature that allows users to find and explore public servers without prior invitation~\cite{discord_discovery}. Accessible via a compass icon at the bottom of the server list (Figure~\ref{fig:screenshot-discover}), the discovery interface presents servers organized by category (Gaming, Music, Education, etc.) and supports keyword search. Servers must meet certain requirements---including minimum member counts, age thresholds, and activity levels---to appear in discovery.

\begin{figure}[t]
    \centering
    \includegraphics[width=0.8\textwidth]{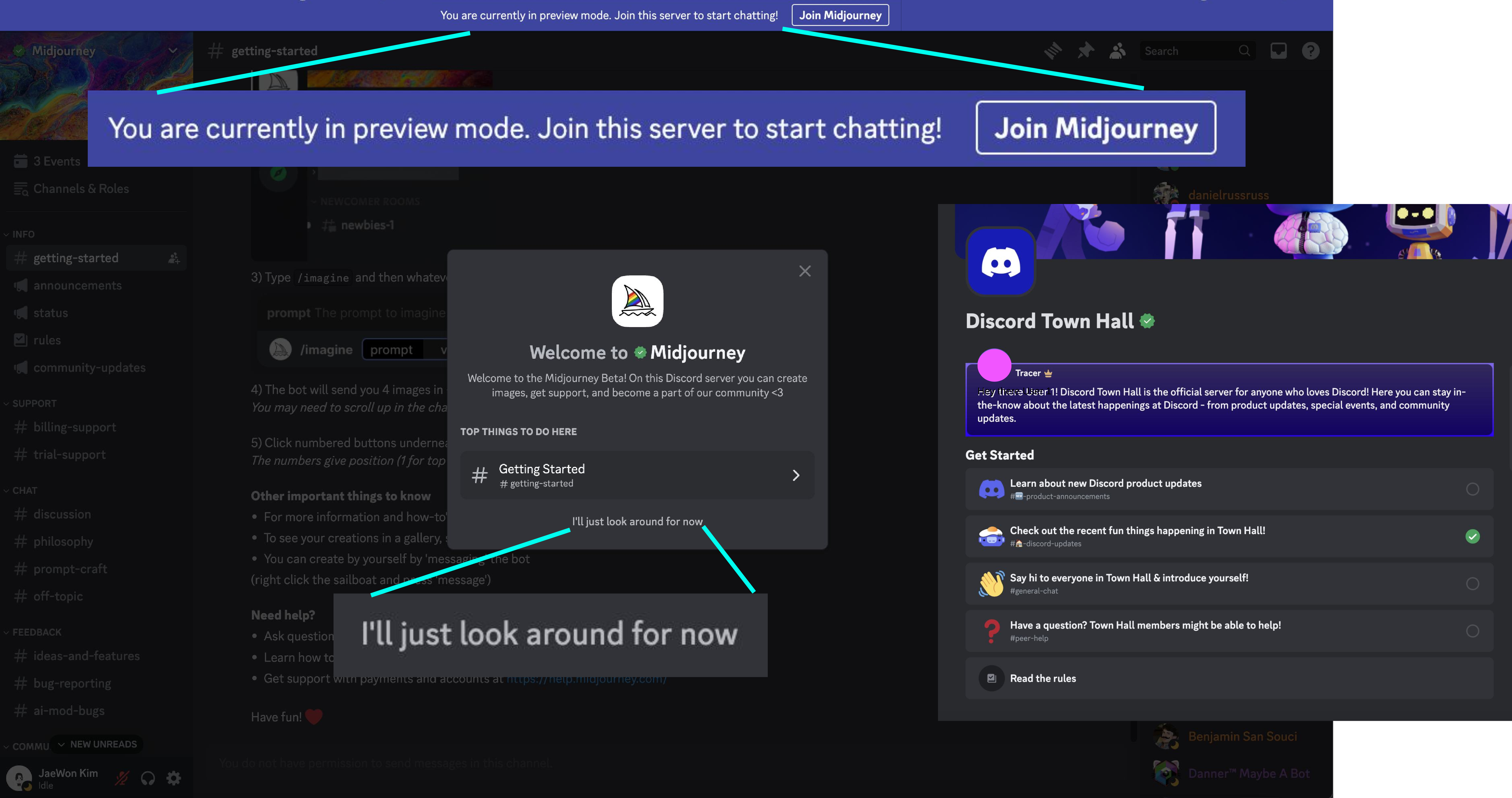}
    \caption{Server preview mode with limited visibility before officially joining.}
    \Description {This image shows a Discord interface with three highlighted sections. At the top, a banner in a blue background displays the text ``You are currently in preview mode. Join this server to start chatting!'' followed by a button labeled ``Join Midjourney.'' In the middle, there is a welcome popup for the ``Midjourney'' server, which has a sailboat icon and the text ``Welcome to Midjourney'' along with a description about creating images, getting support, and being part of the community. Below this description, there is a section labeled ``TOP THINGS TO DO HERE'' with a button titled ``Getting Started'' and a hashtag channel ``\# getting-started.'' At the bottom of this popup, there is an option with the text ``I'll just look around for now'' in a gray box. The background includes various server elements, such as the channel list on the left and a ``Discord Town Hall'' panel on the right.}
    \label{fig:screenshot-visitor}
\end{figure}

Server Discovery includes a ``Preview Mode,'' which allows users to lurk a server's public channels before committing to membership~\cite{discord_discovery}. In this mode, users can read messages and observe activity without officially joining, and the server appears only temporarily in their sidebar (Figure~\ref{fig:screenshot-visitor}). Lurkers' visibility is limited to channels that administrators have marked as public to visitors; private or role-restricted channels remain hidden, and lurkers themselves may not be aware of content they cannot see.

\parHeading{Notification and Presence Controls}
Discord provides granular notification controls that users can configure at multiple levels~\cite{discord_notifications}. At the broadest level, users can set default notification preferences for all servers. Within each server, users can choose to receive notifications for all messages, only messages that mention them directly (via @username or @role), or no notifications at all. Individual channels can override server-wide settings, enabling users to silence busy channels while staying alerted to important announcements; some servers additionally ask new members which topics they want notifications for during onboarding (Figure~\ref{fig:screenshot-notification}). Users can also set their visible presence (online, idle, do not disturb, or invisible), so that reachability itself is adjustable.

\clearpage

%% file: inserts/demographics-individual.tex

\begingroup
\small
\begin{longtable}{@{}l l p{1.55cm} >{\centering\arraybackslash}p{1.2cm} >{\centering\arraybackslash}p{1.2cm} >{\centering\arraybackslash}p{1.2cm} >{\raggedright\arraybackslash}p{5.8cm}@{}}
\caption{Participant characteristics and Discord use. Gender entries reflect participants' selections; P19 and P22 selected two options. Months reports how long each participant had used Discord at the time of the interview. Joined counts all servers they had ever joined; Regular counts those they used regularly. N/A indicates that the participant did not report the information.}
\label{tab:demographics-individual} \\
\toprule
\textbf{ID} & \textbf{Age} & \textbf{Gender} & \multicolumn{3}{c}{\textbf{Discord use}} & \textbf{Uses of Discord} \\
\cmidrule(lr){4-6}
& & & \textbf{Months} & \textbf{Joined} & \textbf{Regular} & \\
\midrule
\endfirsthead
\toprule
\textbf{ID} & \textbf{Age} & \textbf{Gender} & \multicolumn{3}{c}{\textbf{Discord use}} & \textbf{Uses of Discord} \\
\cmidrule(lr){4-6}
& & & \textbf{Months} & \textbf{Joined} & \textbf{Regular} & \\
\midrule
\endhead
\bottomrule
\endfoot
    P01 & 19 & Girl/Woman & 48 & 40 & 1--2 & Work; gaming and talking with friends; following student clubs related to gaming and other topics \\
    P02 & 17 & Boy/Man & 84 & 150+ & 5 & Playing and streaming games; talking with friends; interacting with streamers and fan communities \\
    P03 & 17 & Boy/Man & 36 & 20 & 7 & Talking with friends who play games; moderating; finding scholarships and scholarship communities \\
    P04 & 24 & Boy/Man & 3 & N/A & 2 & Finding local communities based on shared interests \\
    P05 & 19 & Girl/Woman & 53 & N/A & N/A & Joining school servers for study groups and social interaction; joining servers related to games and other interests \\
    P06 & 16 & Boy/Man & 36 & 20--30+ & 4--5 & School; talking with friends; making plans; gaming and general conversation \\
    P07 & 20 & Boy/Man & 62 & 9 & N/A & Talking with friends; managing servers; maintaining a gaming community connected to League of Legends, YouTube, and streaming \\
    P08 & 16 & Girl/Woman & 60 & 20 & 5 & Talking with friends; playing games; accessing servers related to school programs and other interests \\
    P09 & 19 & Girl/Woman & 12 & N/A & N/A & Getting help and meeting people in classes; sharing notes; gaming with friends \\
    P10 & 18 & Non-binary or Third Gender & 24 & 13 & 2 & Following drama club and theater updates; talking with distant friends; joining school and fandom servers, including Dungeons \& Dragons \\
    P11 & 18 & Girl/Woman & 60 & 15--20 & 7 & Group messaging; participating in college student leadership; previously participating in video-game communities \\
    P12 & 19 & Boy/Man & 72 & 40 & 12 & Pursuing hobbies; gaming; interacting with offline friends \\
    P13 & 20 & Girl/Woman & 48 & 27--30 & N/A & Class communication; clubs; servers related to personal interests \\
    P14 & 15 & Girl/Woman & 18 & 12--15 & 6 & Discussing interests such as anime and K-pop; messaging friends; coordinating activities, often through video calls \\
    P15 & 17 & Girl/Woman & 36 & 6 & 4 & Communicating with people who use Discord; coordinating extracurricular activities \\
    P16 & 16 & Girl/Woman & 48 & 240 & 6 & YouTuber communities; school activities; gaming and socializing; screen sharing with a partner \\
    P17 & 16 & Girl/Woman & 24 & N/A & 10 & Communicating within specific communities, a cheerleading team, and school-related servers \\
    P18 & 17 & Girl/Woman & 40 & 15--20 & 2--3 & Playing and streaming video games; participating in group chats \\
    P19 & 20 & Girl/Woman, Non-binary or Third Gender & 30 & 30 & 5 & Playing games and supporting streamers; joining servers related to mental health and other interests \\
    P20 & 19 & Boy/Man & 78 & 15--20 & 5 & Participating in an online academic program; previously gaming \\
    P21 & 15 & Girl/Woman & 36 & 80 & 10 & Following interests such as streaming, the musical Hamilton, and learning French; joining gaming servers \\
    P22 & 19 & Girl/Woman, Non-binary or Third Gender & 100 & 150--200 & 100 & Joining hobby-based servers related to gaming and programming; joining servers with friends \\
    P23 & 18 & Girl/Woman & 55 & 130 & 32 & Communicating with friends; meeting people; playing games \\
    P24 & 18 & Girl/Woman & 20 & 15 & 5--6 & Communicating with friends; participating in disability- and health-related servers; moderating servers \\
    P25 & 19 & Girl/Woman & 74 & 104 & 3+ & Communicating with friends; discussing art and aesthetics; interacting with streamers; moderating \\
\end{longtable}
\endgroup


%% file: inserts/protocol.tex

\section*{Introduction}

Welcome, and thank you for participating in our study. Before we begin, I want to ensure that you've read the consent form I sent earlier. Have you had a chance to review it?

\textbf{[Response]}

Today we’re going to talk about your experiences making new connections or friends over Discord.

The interview will last no longer than an hour, and we will make sure that none of the information that you share in this interview will be shared with anyone in a personally identifiable way.

Please remember that you should feel free to decline to answer a question if you feel uncomfortable with it.

Just to reiterate, this session will be recorded for research purposes. Do I have your consent to proceed with the recording?

\textbf{[Response]}

Thank you. If you have any questions or concerns regarding this interview during or after our interview, please feel free to email me through my email.

\section*{Interview}

I first want to ask you a few general questions about your experience with Discord.

\begin{enumerate}[label=\arabic*.]
    \item Do you remember when you first decided to download and join Discord?
        \begin{enumerate}[label=\alph*.]
            \item How did you learn about the platform and why did you decide to use it?
            \item What is it known for amongst your friends?
            \item What was your first impression of Discord?
        \end{enumerate}

    \item How often do you use it?
        \begin{enumerate}[label=\alph*.]
            \item What do you mainly use it for?
            \item What’s your typical session when you use Discord?
            \item How many servers do you have joined?
            \item Could you briefly describe each server? Maybe, like how many members there are, what the purpose of it is, etc.
        \end{enumerate}

    \item Can you walk me through your typical experience of joining a new Discord server?
    
    \item How do you make sense of a new server? Its rules, its vibes, the people there, and more.

    \item Could you briefly describe how your experience has been like making new connections or friendships over Discord?
        \begin{enumerate}[label=\alph*.]
            \item Who, what kind, where, when, how, why?
            \item Is there a specific reason to choosing Discord to make friends, or no?
        \end{enumerate}
    
    \item Can you recall any of your typical first interactions with strangers on Discord?
        \begin{enumerate}[label=\alph*.]
            \item Where and how did you meet?
            \item How big was the server or the channel that you both were a part of?
            \item What element of Discord do you think helped you the most with that experience of meeting someone, a potential friend, for the first time?
            \item How might other platforms compare to Discord in this process?
        \end{enumerate}

    \item Can you describe your typical experience when you transitioned from being a stranger to becoming friends with someone on Discord?
        \begin{enumerate}[label=\alph*.]
            \item How do you decide who to develop stronger relationships with?
                \begin{enumerate}[label=\roman*.]
                    \item What are some signals that make you decide not to want to develop a connection with someone?
                \end{enumerate}
            \item How did the relationship/friendship progress?
                \begin{enumerate}[label=\roman*.]
                    \item Where and how do you keep in contact with them?
                \end{enumerate}
            \item What element of Discord do you think helped you the most with that experience?
            \item How might other platforms compare to Discord in this process?
        \end{enumerate}

    \item Are there any unique aspects of Discord that you think contribute to friendship development?
        \begin{enumerate}[label=\alph*.]
            \item What features of Discord do you think contribute most to building friendships?
            \item How do you use these to maintain and deepen your friendships?
            \item Are these features available on other platforms too? If so, how do they feel different?
            \item Are there particular channels or servers that helped you with making connections? What were their unique characteristics?
            \item Does the size of the channels or servers affect how you initiate or follow up with other users there?
            \item Did the theme or topic of the server affect your friendship-building?
        \end{enumerate}

    \item What might be some hurdles to making friends over Discord, if any?
    
    \item Which social media platforms, besides Discord, do you use or have used before?
        \begin{enumerate}[label=\alph*.]
            \item Which platforms were the most effective for you in making and/or developing new connections?
            \item How does Discord compare to that platform? What’s unique or different about Discord?
            \item What is your favorite thing about Discord compared to the other platforms?
            \item What is your least favorite thing about Discord compared to the other platforms?
            \item If you could change or add any features to Discord to better support friendship development, what would they be? Why?
            \item If you were to describe the vibe of each platform metaphorically, how would you describe them?
        \end{enumerate}
\end{enumerate}


%% file: inserts/screenshots/screenshot-2.tex
\begin{figure}[htbp]
    \centering
    \includegraphics[width=0.95\linewidth]{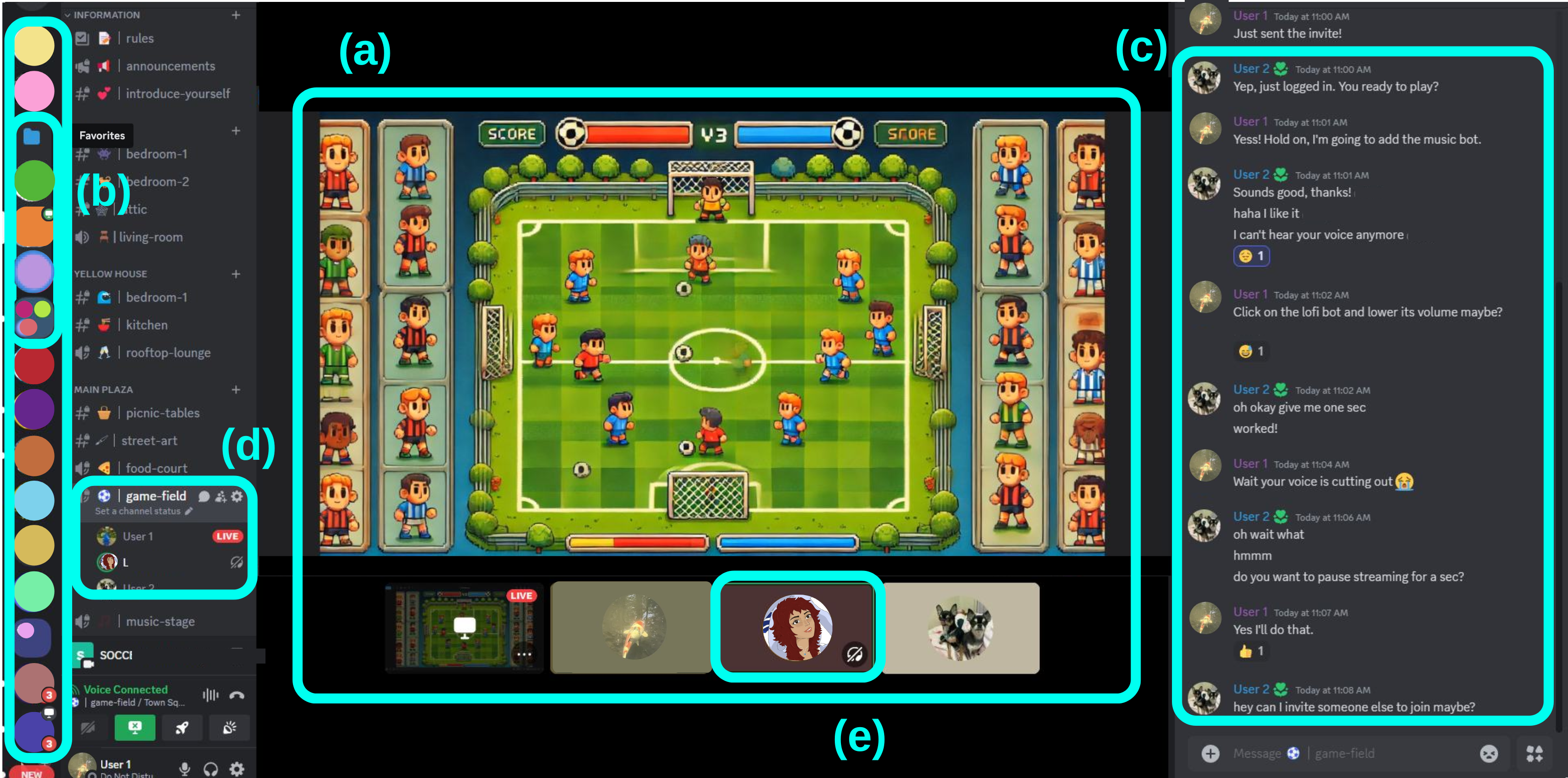}
    \caption{Annotated mockups of Discord features around conversation support: (a) Soccer game streaming in the ``game-field'' voice channel; (b) Re-orderable server list---allows users to organize servers into folders based on custom topics; (c) Text chat window for a voice channel; (d) Public voice channel---users can join by clicking the channel, with visibility of current participants; (e) Music bot---users can activate a bot to play background music for everyone in the channel.}
    \Description {This image is a mock screenshot of a Discord server interface with five highlighted sections for different functions. Section labeled (a): In the center, there is a game in progress with pixel-art graphics, representing a soccer match. Players and avatars are positioned around the field, and the game's visual style suggests a casual, arcade-style soccer game; Section labeled (b): On the left, there is a vertical panel listing various servers, grouped into different folders; Section labeled (c): On the right side of the screen, there is a chat window showing a conversation between two users (User 1 and User 2). The discussion involves troubleshooting audio issues, inviting other users to join, and controlling bots; Section labeled (d): In the channel ``game-field,'' there are users actively participating in a voice channel. Icons indicate that User 1, the bot Lofi, and User 2 are currently live or connected to the ``game-field'' channel. The presence of a small ``LIVE'' badge signals that an activity is being streamed in the voice channel; Section labeled (e): This section illustrates that a bot, in this case Lofi, is currently active in the voice channel, showing that bots, along with users, can join and participate in voice channels.}
    \label{fig:screenshot-2}
\end{figure}

%% file: inserts/screenshots_2.tex
\begin{figure}[t]
\centering
\begin{minipage}[t]{0.3\textwidth}
    \centering
    \includegraphics[width=\textwidth]{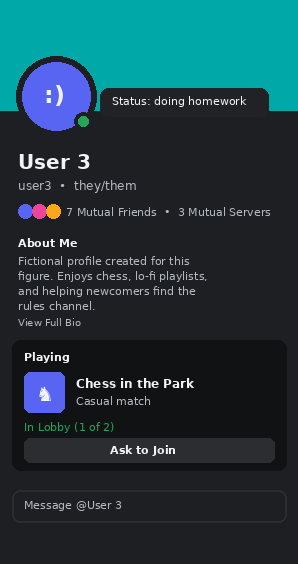}
    \caption{Mock-up of a user profile displaying a status message, current activity, online presence indicator, and mutual-connection information. All profile details are fictional.}
    \Description {This image is a mock-up of a Discord user profile panel for a fictional user named ``User 3.'' The panel background is dark gray with text in white and light gray. At the top, a teal banner sits behind a circular purple avatar with a simple smiling face. To the right of the avatar is a status bubble reading ``Status: doing homework.'' A small green dot on the avatar's edge indicates online status. Below, the display name ``User 3'' appears in bold white text, followed by the handle ``user3'' and pronouns ``they/them.'' Underneath, three small colored circles represent mutual connections next to the text ``7 Mutual Friends and 3 Mutual Servers.'' An ``About Me'' section contains a short fictional bio stating that the profile was created for this figure and mentioning chess, lo-fi playlists, and helping newcomers find the rules channel, with a ``View Full Bio'' link. In the ``Playing'' section, a dark box indicates the user is playing ``Chess in the Park,'' a casual match, with a green line reading ``In Lobby (1 of 2)'' and a gray ``Ask to Join'' button. At the bottom is a ``Message @User 3'' input field.}
    \label{fig:screenshot-status}
\end{minipage}%
\hspace{1em}
\begin{minipage}[t]{0.55\textwidth}
    \centering
    \includegraphics[width=\textwidth]{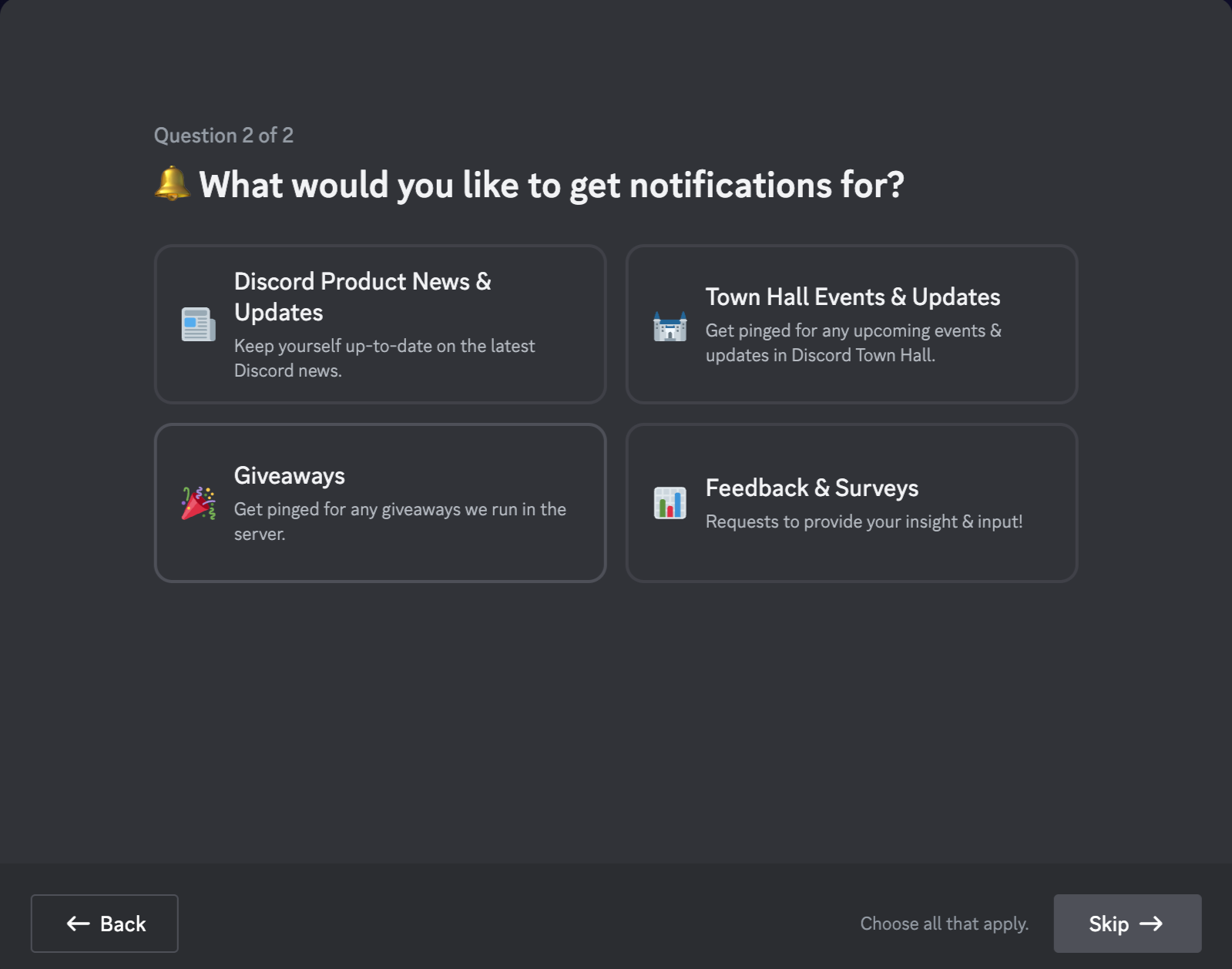}
    \caption{Onboarding screen from Discord's official Town Hall server, in which new members choose which topics they want to be notified about. Beyond such server-configured prompts, users can adjust or mute notifications for each server and channel.}
    \Description {This image displays a screen titled ``Question 2 of 2'' with a bell emoji followed by the heading: ``What would you like to get notifications for?'' There are four rectangular options on the screen, arranged in a 2x2 grid, each representing a different notification category. In the top left corner, the first option is titled ``Discord Product News \& Updates,'' with an icon of a newspaper beside it. The description under the title reads: ``Keep yourself up-to-date on the latest Discord news.'' The top right corner contains the second option, titled ``Town Hall Events \& Updates,'' accompanied by an icon of a blue castle. The description states: ``Get pinged for any upcoming events \& updates in Discord Town Hall.'' In the bottom left corner, the third option is labeled ``Giveaways,'' with a party popper emoji icon next to it. The description says: ``Get pinged for any giveaways we run in the server.'' The bottom right corner shows the fourth option, ``Feedback \& Surveys,'' with an icon of a bar chart. The description reads: ``Requests to provide your insight \& input!'' At the bottom of the screen, there is a left-pointing arrow labeled ``Back'' on the left side. On the right, there is a ``Skip'' button with a right-pointing arrow. Below the notification options, a note reads: ``Choose all that apply.'' The interface uses a dark background with light text and icons.}
    \label{fig:screenshot-notification}
\end{minipage}

\vspace{1em}

\begin{minipage}[t]{\textwidth}
    \centering
    \includegraphics[width=0.75\textwidth]{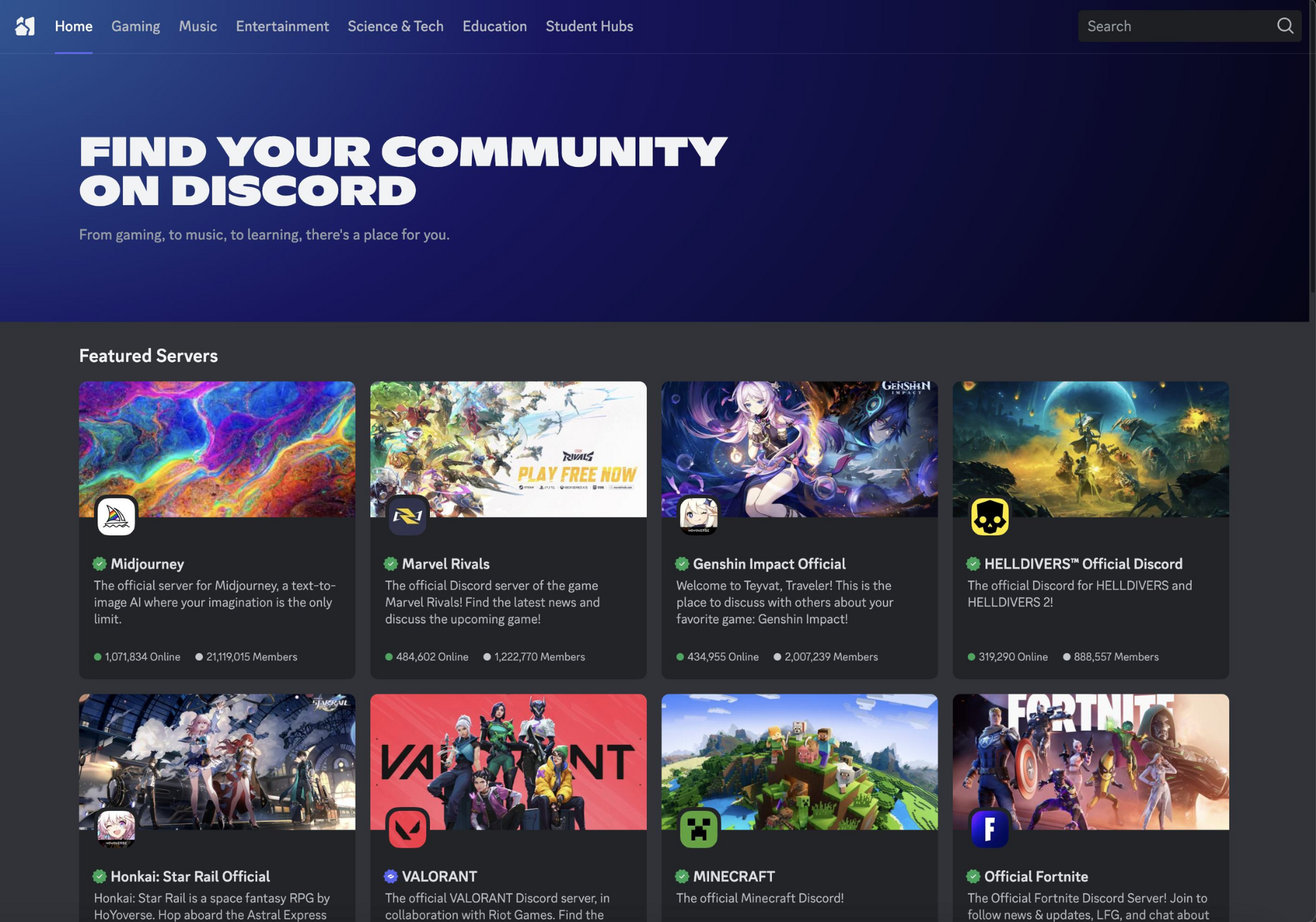}
    \caption{Server Discovery interface allowing users to browse public servers by category and keyword search.}
    \Description {This image shows the Discord discover page for finding servers. At the top, there is a large blue banner with white text that reads ``FIND YOUR COMMUNITY ON DISCORD'' and a subheading that says, ``From gaming, to music, to learning, there's a place for you.'' Below the banner, the page displays a grid of server thumbnails under the heading ``Featured Servers.'' Each thumbnail represents a different community, featuring categories like gaming, AI, and entertainment. Each server card includes the server’s name, an icon, a brief description, and information about the number of online users and total members. Some of the servers displayed include ``Midjourney,'' ``Marvel Rivals,'' ``Genshin Impact Official,'' ``HELLDIVERS\texttrademark{} Official Discord,'' and others. The interface has a search bar in the top right corner and navigation options at the top for browsing different categories, including ``Home,'' ``Gaming,'' ``Music,'' ``Entertainment,'' ``Science \& Tech,'' ``Education,'' and ``Student Hubs.''}
    \label{fig:screenshot-discover}
\end{minipage}
\end{figure}